\def\be{\begin{equation}}
	\def\ee{\end{equation}}
\def\ba{\begin{array}}
	\def\ea{\end{array}}

\documentclass[prl,showpacs,twocolumn,amsmath]{revtex4}
\usepackage{amsmath}
\usepackage{amsfonts}
\usepackage{mathrsfs}
\usepackage{amssymb}
\usepackage{pifont}
\usepackage{epsfig,subfigure,dsfont,amsthm,amsbsy,mathrsfs,amscd}
\usepackage{epstopdf}
\usepackage{bbm}
\usepackage{color}

\def\qed{\leavevmode\unskip\penalty9999 \hbox{}\nobreak\hfil
	\quad\hbox{\leavevmode  \hbox to.77778em{%
			\hfil\vrule   \vbox to.675em%
			{\hrule width.6em\vfil\hrule}\vrule\hfil}}
	\par\vskip3pt}
\usepackage{leftidx}

\input amssym.def
\begin{document}
	\title{\large\bf  Dynamics of Entanglement in Schwarzschild Black Holes}
	
	\author{Fang Xie$^{1, ^\ast}$, Ying Yang$^{1, ^\ast}$, Tinggui Zhang$^{1}$, Xiaofen Huang$^{1, ^\dag}$}
	\affiliation{ School of Mathematics and Statistics, Hainan Normal University, Haikou, 571158, China \\
		$\ast$ These authors contributed equally to this work \\
        $^{\dag}$ Correspondence to  huangxf1206@163.com}
	
	\bigskip
	\bigskip

\begin{abstract}
To characterize the effect of Hawking radiation induced by the quantum atmosphere beyond the event horizon on entanglement, we employ concurrence as the entanglement measure for a bipartite mixed state and investigate its evolution with Hawking temperature. We find that the physically accessible concurrence decreases as the Hawking acceleration increases, whereas the physically inaccessible concurrence exhibits the opposite behavior, increasing monotonically from zero. We further establish several trade-off relations on concurrence, revealing its distribution between physically accessible and inaccessible regions. Additionally, we study the dynamics of concurrence under three types of channel noise. The results indicate that the evolution of concurrence depends on the specific noise channel: unlike the phase damping channel, sudden death of concurrence occurs in both phase flip and bit flip channels, the concurrence exhibits a certain symmetry with respect to the noise parameter during its evolution under bit flip channel noise.

\end{abstract}

 \maketitle

\section{I. Introduction}
Within the framework of Einstein general relativity, the gravitational collapse of sufficiently massive stars is predicted to form black holes in our universe. Recent observations of black hole imaging and gravitational wave detections have significantly advanced our understanding of black hole formation, evolution, and the behavior of strong-field gravity \cite{hawking1975, hawking1976, h3, h4, bhole3}. Despite considerable progress, the intrinsic properties of black holes and their vast distances from Earth continue to pose substantial challenges, leaving many fundamental mysteries unresolved. Consequently, research on black holes remains largely theoretical to date. From a classical perspective, once a particle crosses a black hole event horizon, it cannot return or escape, as the horizon acts as a one-way boundary.
However, given quantum effects, the particles inside the black hole are destined to gradually escape to the outside, resulting in Hawking radiation \cite{hawking, hawking1975, hawking1976}. This phenomenon is an intermediate bridge between quantum mechanics and gravity, and is at the heart of the information paradox of black hole \cite{bhole4, hawking1975, hawking1976}. Recently, it has made significant progress in research on understanding of quantum phenomena in black hole setting, numerous studies have demonstrated that quantum correlations such as entanglement, coherence, and uncertainty relations can be degraded by Hawking radiation in relativistic settings \cite{entang1, entang2, entang3, entang4, entang5, cohen1, wd2024, wd2025,  Mgw2024, Mgw2025, Ztg2023, con, wu2019, wu2021, wz2021, har2021, unrh1, unrh2, unrh3, unrh4}.

Entanglement, which reveals the intrinsic nature of quantum mechanics, serves as a key distinction between quantum and classical mechanics. As a fundamental resource in quantum processing, it plays a significant role in quantum communication\cite{app1, app2}, and quantum cryptography cryptographic protocols \cite{app3}.  For both theoretical and potentially
 practical reasons, it has become interesting to quantify entanglement, just as
 we quantify other resources such as energy and information.
For the bipartite system, various
entanglement measures have been proposed, such as concurrence \cite{woot97}, negativity \cite{werner}, entanglement of formation \cite{woot98}. In particular,  Carvalho \textit{et al}. derived the generalization of concurrence, namely the generalized multipartite concurrence \cite{gcon}.

In order to better understand quantum properties of the black hole, examining entanglement and quantumness in Schwarzschild spacetime is required. In particular, quantum entanglement influenced by Hawking radiation is a possible important way to solve the information paradox of the black hole.
In Ref. \cite{qiang2018}, the authors investigated the Unruh effect on the genuine concurrence for Greenberger-Horne-Zeilinger-like state in noninertial frames.
In Ref. \cite{wsm2024}, S. Wu \textit{et al}. studied the genuine concurrence for the N-partite Greenberger-Horne-Zeilinger states.  In Ref. \cite{kim}, K. Kim \textit{et al.} considered the entanglement of the tripartite $W$ state via $\pi$-tangle. These studies have focused on the evolution of entanglement in noninertial frames, yet they have predominantly considered pure states as the initial states. The behavior of entanglement evolution for mixed states remains unclear.

In our work, we investigate mixed states as initial configurations and analyze the evolution of their entanglement under Hawking radiation. This study serves as a valuable complement to earlier research.
 We consider Alice and Bob sharing an isotropic mixed state in Minkowski space, with both observers accelerating near the event horizon. We will focus on studying the influence of Hawking radiation and channel noise on concurrence, deriving analytical expressions for concurrence in physical accessible and physical inaccessible modes, and analyzing concurrence dynamics under phase damping, phase flip and bit flip channels.
This study establishes a nexus between relativistic quantum information and open quantum systems, elucidating the cooperative regulation of entanglement by environmental noise and spacetime curvature a pivotal advancement toward realizing operational quantum protocols in relativistic frameworks.

The structure of the paper is as follows. In Sec. II, we introduce the quantization of  Schwarzschild spacetime. In Sec. III, we discuss concurrence of two-qubit
mixed states in the Schwarzschild black hole. In Sec. IV, we investigate the concurrence under both noisy environment and Unruh effect. The last section is
devoted to the summary.

\section{II.  The mathematical characterization of the Dirac field in Schwarzschild spacetime}
First, we provide a brief overview of the vacuum state in a Schwarzschild black hole. The Schwarzschild black hole is the simplest black hole solution in general relativity---non-rotating, uncharged, and spherically symmetric.
The spacetime geometric outside the Schwarzschild black hole can be expressed as
\begin{equation}\label{R1}
ds^2=(1-\frac{2M}{r})^{-1}dr^2-(1-\frac{2M}{r})dt^2+r^2(d\theta^2+\sin \theta^2d\phi^2),
\end{equation}
where $M$ is the mass of the black hole, $r$ is the radial distance.

The Dirac's equation in Schwarzchild spacetime is given by
$
[\gamma^{\alpha}e_{\alpha}^{\mu}(\alpha_{\mu}+\Gamma_{\mu})]\psi=0,
$
where $\gamma^{\alpha}$ stands for the Dirac' matrices and $\Gamma_{\mu}$ is the spin connection coefficient, which has a set of solutions for positive frequency outgoing solutions for the inside and
outside regions of the event horizon (Schwarzschild modes), that is,
\begin{equation}\label{R2}
\left\{
  \begin{split}
   \psi_k^{I+}&=\varsigma(r) e^{-\textrm{i}\omega\mu}, &   \\
    \psi_k^{II+}&=\varsigma(r) e^{\textrm{i}\omega\mu}, &
  \end{split}
\right.
\end{equation}
where $\varsigma(r)$ stands for the four-component Dirac spinor, and $\mu=t-r_{*}$ with the tortoise coordinate
$r_{*}=r+2M \textrm{ln} \frac{r-2M}{2M}$.

By taking the above two positive-frequency wave solutions as a set of completely orthogonal basis, and expanding the Dirac field, one can obtain
\begin{equation}\label{R3}
\psi_{out}=\sum_{\chi=I, II}\int dk (\alpha_k^{\chi}\psi_k^{\chi^{+}})(\beta_k^{\chi^{*}}\psi_k^{\chi^{-}}),
\end{equation}
where $\alpha$ and $\beta$ are the fermion's annihilation operator and antifermion's creation operator,  respectively.

According to the relation between Kruskal coordinates and black hole coordinates, a new set of
orthogonal basis can be obtained (Kruskal modes), which is a complete basis for the positive
energy mode,
\begin{equation*}\label{R4}
\left\{
  \begin{array}{ll}
   \phi_k^{I+}=e^{2\pi M\omega_k}\psi_k^{I^{+}}+e^{-2\pi M\omega_k}\psi_{-k}^{II^{-}}, & \hbox{}  \\
    \phi_k^{II+}= e^{-2\pi M\omega_k}\psi_{-k}^{I^{-}}+e^{2\pi M\omega_k}\psi_k^{II^{+}}. & \hbox{}
  \end{array}
\right.
\end{equation*}

In this way, the new representation of the Dirac field in Kruskal spacetime can be expressed as
\begin{equation}\label{R5}
\psi_{out}=\sum_{\chi=I, II}\int dk \frac{1}{\sqrt{2cosh(4\pi M \omega_k)}}(a_k^{\chi}\phi_k^{\chi^{+}}+b_k^{\chi^{*}}\phi_k^{\chi^{-}}),
\end{equation}
where $a_k^{\chi}$ and $b_k^{\chi^{*}}$ are the fermion annihilation and antifermion creation operators acting on the Kruskal vacuum respectively.
It means the Dirac field can be quantized by both Schawrzschild and Kruskal modes.

By Bogoliubov transform, based on (\ref{R3}) and (\ref{R5}), we can get the annihilation operator in the following form
\begin{equation}\label{R6}
c_k^I=\cos ra_k^I-\sin r\beta_k^{II*},
\end{equation}
where $r$ is acceleration parameter, $\cos r=(e^{-\frac{\omega}{T}}+1)^{-\frac{1}{2}}$,  $\sin r=(e^{\frac{\omega}{T}}+1)^{-\frac{1}{2}}$, $\omega$ stands for the
monochromatic frequency, and $T=\frac{1}{8\pi M}$ denotes the Hawking temperature.

What is more, we can obtain the vacuum state and the  excited state of the Kruskal mode in Schwarzschild spacetime,
\begin{equation}\label{R7}
\begin{aligned}
  |0\rangle \rightarrow   &  \cos r|0\rangle_{I}|0\rangle_{II}+\sin r|1\rangle_{I}|1\rangle_{II},  \\
  |1\rangle  \rightarrow  &  |1\rangle_{I}|0\rangle_{II}.
\end{aligned}
\end{equation}
Herein, $\{ |n\rangle_{I(II)}\} $ are the orthogonal bases for the outside region and inside region of the event horizon, respectively.

\section{III. Concurrence for two-qubit states in Schwarzchild spacetime}
In this section, we adopt the concurrence of a two-qubit quantum system as the entanglement measure, focusing on the influence of Hawking radiation on entanglement. The temporal evolution of entanglement is analyzed numerically.

\subsection{A. Concurrence for two-qubit states}

Concurrence serves as an important measure of entanglement \cite{con1, con2, con3, con4}. For pure states, it possesses an explicit and concise computational formula $C(\psi)=\sqrt{2(1-\textmd{Tr}\rho^2_A)}$, where $\rho_A=\textmd{Tr}_B (\rho_{AB})$ is the reduced density matrix. However, for mixed states, calculating concurrence becomes considerably more challenging, which is given by the convex roof construction, the minimum average concurrence taken over all ensemble decompositions of $\rho_{AB}$,
\begin{equation*}
C(\rho)=\min\limits_{\{(p_i, \psi_i)\}}\sum_i p_iC(\psi_i).
\end{equation*}

Nevertheless, Wootters \cite{woot98} established a formulation of entanglement quantification via concurrence, enabling precise evaluation of entanglement for arbitrary two-qubit systems. For two-qubit quantum states, concurrence can be computed via the Bloch representation of the quantum state. Recall the Bloch representation for a two-qubit state $\rho$, it states that there is a decomposition according to the Pauli matrices, that is,
\begin{equation}\label{R8}
\begin{aligned}
\rho=\frac{1}{4}(I\otimes I+\mathbf{r}\cdot\boldsymbol{\sigma}\otimes I + \mathbf{s}\cdot I\otimes\cdot\boldsymbol{\sigma}+\sum_{i = 1}^{3}c_{i}\sigma_{i}\otimes\sigma_{i}),
\end{aligned}
\end{equation}
where $\mathbf{r}$ and $\mathbf{s}$ are Bloch vectors, $\boldsymbol{\sigma}=(\sigma_1, \sigma_2, \sigma_3)$ and $\{\sigma_{i}\}_{i = 1}^{3}$ are the standard Pauli matrices. When Bloch vectors $\mathbf{r}=\mathbf{s} = 0$, $\rho$ reduces to a two-qubit Bell-diagonal state.
Now assume that the Bloch vectors are $z$ - directional, that is, $\mathbf{r}=(0,0,r)$, $\mathbf{s}=(0,0,s)$.
Exactly, one can always change them to be $x$ or $y$-directional via an appropriate local unitary transformation without losing its diagonal property of the correlation term.
In this case arbitrary state $\rho$ defined in Eq.(\ref{R8}) has a matrix form as follows

\begin{widetext}
\begin{equation}\label{R9}
\begin{aligned}
&\rho =\frac{1}{4}
&\begin{pmatrix}
1 + r + s + c_3 & 0 & 0 & c_1 - c_2 \\
0 & 1 + r - s - c_3 & c_1 + c_2 & 0 \\
0 & c_1 + c_2 & 1 - r + s - c_3 & 0 \\
c_1 - c_2 & 0 & 0 & 1 - r - s + c_3
\end{pmatrix}.
\end{aligned}
\end{equation}
\end{widetext}

 This indicates that the $z$-directional state is an X-state, and the concurrence of such two-qubit X-states can be calculated directly from their density matrices \cite{con}. Denote $\tilde{\rho}= \sigma_y \otimes \sigma_y\rho^{*}\sigma_y \otimes \sigma_y$, and $*$ indicates complex conjugate.
By calculating,  we can obtain four eigenvalues of $\rho\tilde{\rho}$,
\begin{equation}\label{R11}
\begin{aligned}
\lambda_1 &=& \frac{1}{16}\Big(c_1 - c_2 - \sqrt{(1 + r + s + c_3)(1 - r - s + c_3)}\Big)^2, \\
\lambda_2 &=& \frac{1}{16}\Big(c_1 - c_2 + \sqrt{(1 + r + s + c_3)(1 - r - s + c_3)}\Big)^2, \\
\lambda_3 &=& \frac{1}{16}\Big(c_1 + c_2 - \sqrt{(1 + r - s - c_3)(1 - r + s - c_3)}\Big)^2, \\
\lambda_4 &=& \frac{1}{16}\Big(c_1 + c_2 + \sqrt{(1 + r - s - c_3)(1 - r + s - c_3)}\Big)^2.
\end{aligned}
\end{equation}
Then, the concurrence for a two-qubit quantum system with density matrix given in (\ref{R9}) is completely determined by the eigenvalues of $\rho\tilde{\rho}$,
\begin{equation}\label{R12}
\begin{aligned}
C(\rho) =&\max \Big\{ 2 \max \{\sqrt{\lambda_1}, \sqrt{\lambda_2}, \sqrt{\lambda_3}, \sqrt{\lambda_4} \} - \sqrt{\lambda_1} \\
&- \sqrt{\lambda_2}- \sqrt{\lambda_3} - \sqrt{\lambda_4}, 0 \Big\}.
\end{aligned}
\end{equation}
In such case, concurrence serves as a computable entanglement measure, and free entanglement in a quantum state implies that its concurrence vanishes.

\subsection{B.  Concurrence under the Schwarzchild  black hole}
In this subsection, we investigate  the effect of Hawking radiation on concurrence. Now, we assume Alice and Bob share an isotropic state $\rho = \frac{1-p}{4} I \otimes I + p |\psi^+\rangle\langle\psi^+|$ \cite{horo},
where $p\in [0, 1]$, and $|\psi^+\rangle=\frac{1}{\sqrt{2}}(|00\rangle+|11\rangle)$ is the maximally entangled state. It has Bloch decomposition as below
\begin{equation}\label{R13}
\rho=\frac{1}{4}(I\otimes I+ p\sigma_1\otimes\sigma_1+p\sigma_2\otimes\sigma_2-p\sigma_3\otimes\sigma_3).
\end{equation}
 It is clear to know that $\rho$ is an X-state. Furthermore, we assume both Alice and Bob hover near the event horizon of a Schwarzschild black hole with acceleration $r_a$ and $r_b$, respectively. Due to the Hawking radiation, the Dirac fields will change from the perspective of the uniformly accelerating observer. Consequently, the state $\rho$ will evolves into a four-qubit quantum state $\rho_{A_IA_{II}B_IB_{II}}$, whose analytical expression can be derived using Eqs. (\ref{R7}), however, we omit it here due to its complexity.

Since the interior region of a Schwarzschild black hole is causally disconnected from the exterior, we refer to the modes inside the event horizon namely, $A_{II}$ and $B_{II}$ as the inaccessible modes, and those outside the event horizon ($A_I$ and $B_I$) as the accessible modes. As Alice and Bob cannot access region $II$, it is necessary to trace over modes $A_{II}$ and $B_{II}$, resulting in the bipartite mixed state $\rho_{A_I B_I}$ obtained by partial tracing. Its explicit form is given by:
\begin{equation}\label{R14}
\begin{aligned}
\rho_{A_IB_I}=&\frac{1}{4}\Big(I\otimes I-\sin^2 r_a \sigma_3\otimes I-\sin^2 r_b I\otimes \sigma_3\\
&+p\cos r_a\cos r_b\sigma_1\otimes \sigma_1
+p\cos r_a\cos r_b\sigma_2\otimes \sigma_2\\
&+(\sin^2 r_a\sin^2 r_b-p\cos^2 r_a\cos^2 r_b)\sigma_3\otimes \sigma_3\Big).
\end{aligned}
\end{equation}

In an analogous manner, we compute the remaining three reduced density operators,
\begin{equation}\label{R16}
\begin{aligned}
\rho_{A_{II}B_{II}}=&\frac{1}{4}\Big(I\otimes I+\cos^2 r_a \sigma_3\otimes I+\cos^2 r_b I\otimes \sigma_3\\
&+p\sin r_a\sin r_b\sigma_1\otimes \sigma_1+p\sin r_a\sin r_b\sigma_2\otimes \sigma_2\\
&+(\cos^2 r_a\cos^2 r_b-p\sin^2r_a\sin^2 r_b)\sigma_3\otimes \sigma_3\Big),
\\
\end{aligned}
\end{equation}

\begin{equation}\label{R16}
\begin{aligned}
\rho_{A_IB_{II}}=&\frac{1}{4}\Big(I\otimes I-\sin^2 r_a \sigma_3\otimes I+\cos^2 r_b I\otimes \sigma_3\\
&+p\cos r_a \sin r_b\sigma_1\otimes \sigma_1-p\cos r_a\sin r_b\sigma_2\otimes \sigma_2\\
&-(\sin^2 r_a\cos^2 r_b-p\cos^2r_a\sin^2 r_b)\sigma_3\otimes \sigma_3\Big),
\\
\end{aligned}
\end{equation}

\begin{equation}\label{R16}
\begin{aligned}
\rho_{A_{II}B_{I}}=&\frac{1}{4}\Big(I\otimes I+\cos^2 r_a \sigma_3\otimes I-\sin^2 r_b I\otimes \sigma_3\\
&+p\sin r_a \cos r_b\sigma_1\otimes \sigma_1-p\sin r_a\cos r_b\sigma_2\otimes \sigma_2\\
&-(\cos^2 r_a\sin^2 r_b-p\sin^2r_a\cos^2 r_b)\sigma_3\otimes \sigma_3\Big).
\end{aligned}
\end{equation}

Their quantum concurrences, computed using Eq. (\ref{R12}), are given in the following,
\begin{equation}
\begin{aligned}
C&(\rho _{A_IB_I} )=\frac{1}{2}( 2p\cos r_a\cos r_b - \cos r_a\cos r_b\\
&\sqrt{(1-p)( 2\sin^2 r_a + 2\sin^2 r_b + (1-p)\cos^2 r_a\cos^2 r_b )} ),
\end{aligned}
\end{equation}

\begin{equation}
\begin{aligned}
C&(\rho _{A_{II}B_{II}} )=\frac{1}{2}( 2p\sin r_a\sin r_b - \sin r_a\sin r_b\\&\sqrt{(1-p)( 2\cos^2 r_a + 2\cos^2 r_b + (1-p)\sin^2 r_a\sin^2 r_b )} ),
\end{aligned}
\end{equation}

\begin{equation}
\begin{aligned}
C&(\rho _{A_{I}B_{II}} )=\frac{1}{2}( 2p\cos r_a\sin r_b - \cos r_a\sin r_b\\&\sqrt{(1-p)( 2\sin^2 r_a + 2\cos^2 r_b + (1-p)\cos^2 r_a\sin^2 r_b )} ),\\
\end{aligned}
\end{equation}

\begin{equation}
\begin{aligned}
C&(\rho _{A_{II}B_{I}} )=\frac{1}{2}( 2p\sin r_a\cos r_b - \sin r_a\cos r_b\\&\sqrt{(1-p)( 2\cos^2 r_a + 2\sin^2 r_b + (1-p)\sin^2 r_a\cos^2 r_b )} ).
\end{aligned}
\end{equation}
%

When $p=1$, the isotropic state reduces to a maximally entangled pure state, leading to a more concise expression for the concurrence of its reduced states,
\begin{equation}\label{R17}
\begin{aligned}
&C(\rho _{A_IB_I} )=\cos r_a\cos r_b, \\
&C(\rho _{A_{II}B_{II}} )=\sin r_a\sin r_b,\\
&C(\rho _{A_{I}B_{II}} )=\cos r_a\sin r_b,\\
&C(\rho _{A_{II}B_{I}} )=\sin r_a\cos r_b.
\end{aligned}
\end{equation}

These analytical expressions (\ref{R17}), provide the concurrence of bipartite quantum states across different partitions of accelerating observers for a specific parameter configuration. These values depend explicitly on the acceleration parameters
$r_a$ and $r_b$ of the two observers. Clearly, the concurrences are influenced by the Hawking acceleration. An inherent constraint exists between the concurrences in the physically accessible and inaccessible regions, as captured by the following trade-off relation:
\begin{equation}\label{R18}
\begin{aligned}
C(\rho _{A_IB_I} )^2&+C(\rho _{A_{II}B_{II}} )^2+C(\rho _{A_{I}B_{II}} )^2\\
&+C(\rho _{A_{II}B_{I}} )^2=1.
\end{aligned}
\end{equation}
As acceleration increases, the distribution of quantum correlations across different observer partitions evolves. A notable phenomenon is the strict complementary relationship among these four concurrences, reflecting the conservation of quantum resources under partitioned observations. This complementarity is precisely quantified by the constant sum of their squares, which equals 1.

Furthermore, suppose Alice and Bob move to the black hole with the same Hawking acceleration $r$, that is, $r_a=r_b=r$, Eqs. (\ref{R17}) reduce to $C(\rho _{A_IB_I} )= \cos^2r$ and $C(\rho _{A_{II}B_{II}} )=\sin^2r$,
Thus, we establish a trade-off relation between $C(\rho _{A_IB_I} )$ and $C(\rho _{A_{II}B_{II}} )$,
\begin{equation}\label{R181}
C(\rho _{A_IB_I} )+C(\rho _{A_{II}B_{II}} )=1.
\end{equation}
Moreover, if one observer (e.g., Bob) is in the asymptotically flat region (i.e., $r_b = 0$), the concurrences $C(\rho_{A_{II}B_{II}})$ and $C(\rho_{A_{I} B_{II}})$, associated with the physically inaccessible region of $B_{II}$, vanish. In this case, the physically accessible concurrences
$C(\rho_{A_I B_I}) = \cos r_a$ and $C(\rho_{A_I B_{II}}) = \sin r_a$, which
depend entirely on Alice's   acceleration parameter $r_a$.
 Therefore, we derive another trade-off relation,
\begin{equation}\label{R182}
C(\rho _{A_IB_I} )^2+C(\rho _{A_{II}B_{I}} )^2=1.
\end{equation}
Clearly, Eq. (\ref{R182}) represents a special case of Eq. (\ref{R18}), reducing to it when only one observer undergoes acceleration.

Exactly, Eqs. (\ref{R181}) and (\ref{R182}) reflect two interesting phenomenas. On the one hand, it reveals a constraint between physically accessible and inaccessible entanglement: when the reduced state $\rho _{A_IB_I}$ of the physically accessible part is separable, the physically inaccessible entanglement reaches its maximum value of 1. On the other hand, it indicates the range of concurrence for physically accessible and inaccessible entanglement, which spans a minimum of 0 to a maximum of 1.

\begin{figure*}[htbp]
	   	\centering
\vspace{-3.5cm}
		\includegraphics[width=0.4\textwidth]{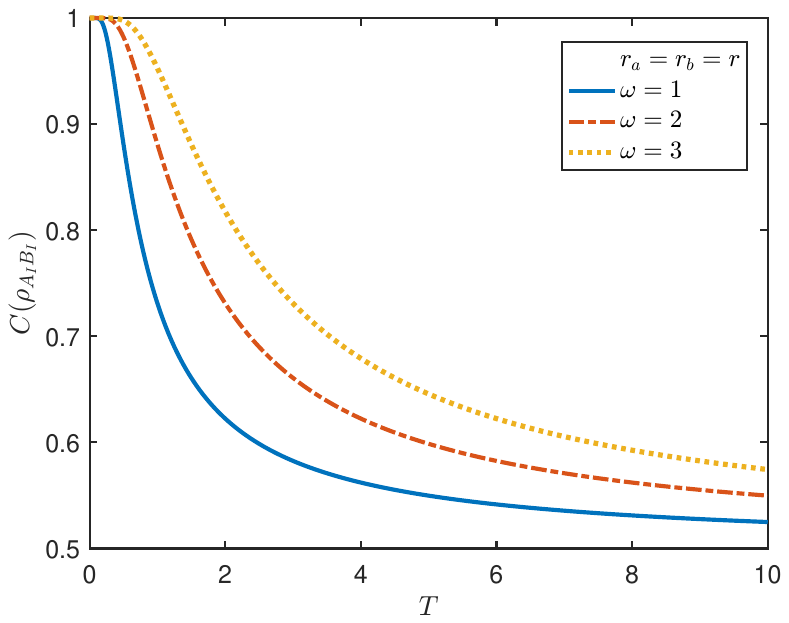}
\hspace{-2cm}
		\includegraphics[width=0.4\textwidth]{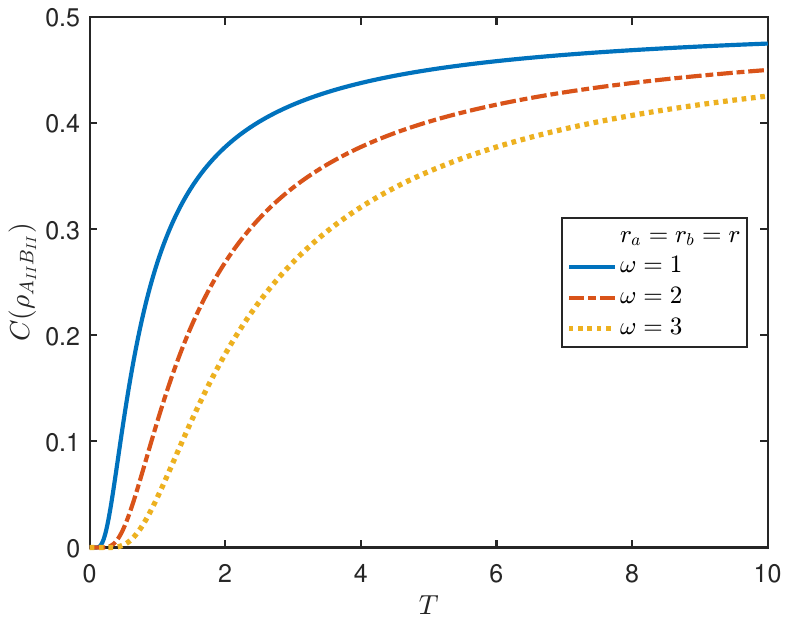}
\hspace{-2cm}
		\includegraphics[width=0.4\textwidth]{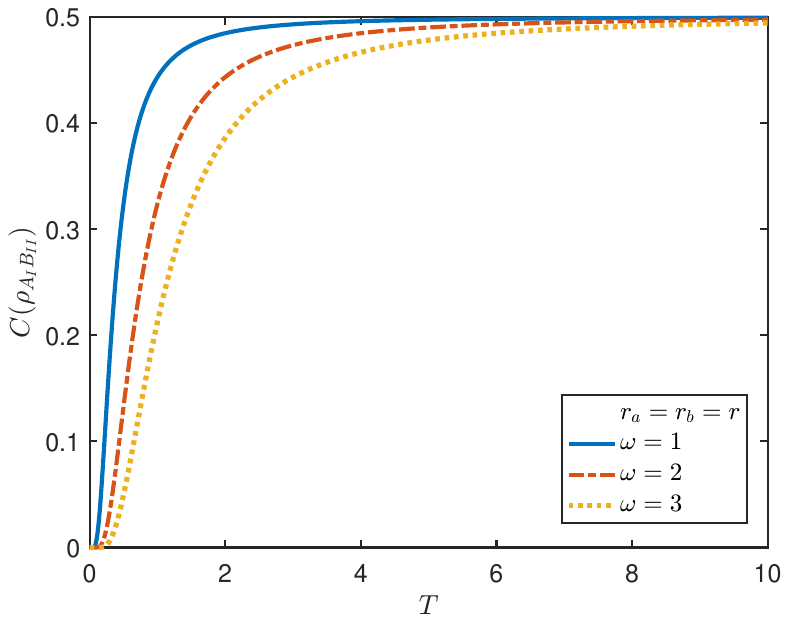}
\vspace{-3.5cm}
\caption{These panels show the evolutions of the concurrences $C(\rho_{A_IB_I})$, $C(\rho_{A_{II}B_{II}})$ and $C(\rho_{A_{I}B_{II}})$ over Hawking temperature. Here we assume Alice and Bob share the same acceleration, and state parameter $p=1$.}
\label{fig:concurrence_plots1}
	   \end{figure*}

\begin{figure*}[htbp]
	   	\centering
\vspace{-3.5cm}
		\includegraphics[width=0.4\textwidth]{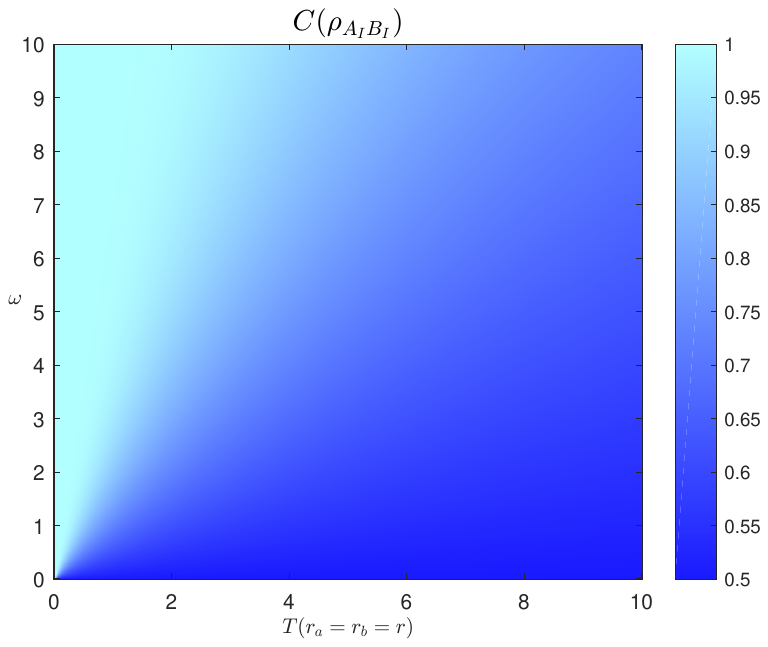}
\hspace{-2cm}
		\includegraphics[width=0.4\textwidth]{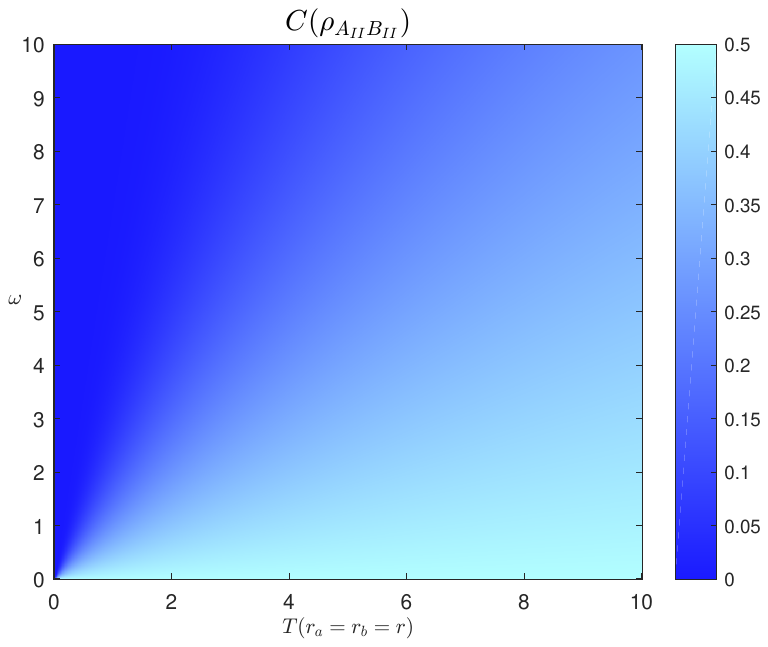}
\hspace{-2cm}
		\includegraphics[width=0.4\textwidth]{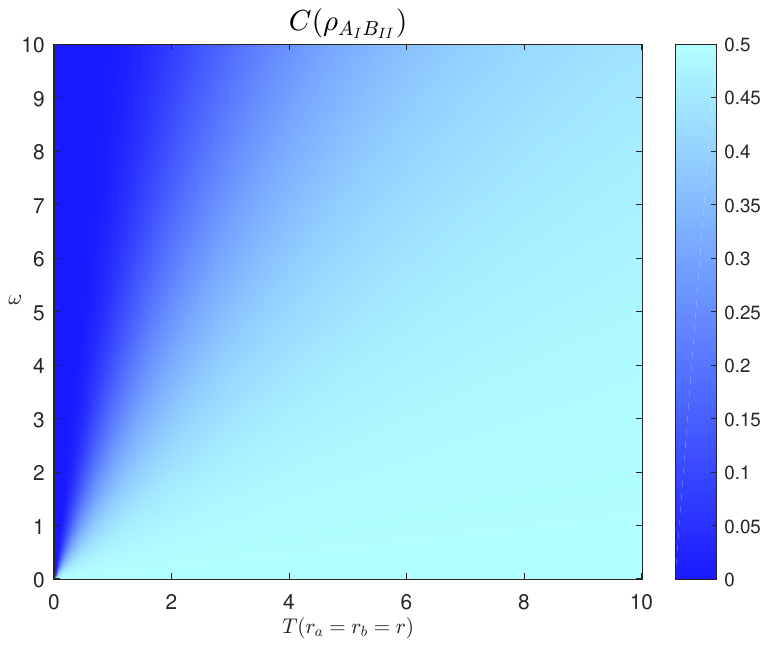}
\vspace{-3.5cm}
\caption{Plot concurrences $C(\rho_{A_IB_I})$, $C(\rho_{A_{II}B_{II}})$ and $C(\rho_{A_{I}B_{II}})$ as functions of $\omega$  and $T$ in the case of Alice and Bob with the same Hawking acceleration and $p=1$.} \label{fig:concurrence_plots2}
	   \end{figure*}

These analytical expressions in Eqs. (\ref{R17}) give the concurrence of bipartite quantum states between different partitions of accelerating observers under the specific parameter setting.
These concurrence values depend explicitly on the acceleration parameters $r_a$ and $r_b$ of the two observers.
To gain a deeper understanding,
we take  concurrence as a function of Hawking temperature $T$ and monochromatic frequency $\omega$. We plot concurrences of bipartite reduce states with various values for parameters.
As shown in Fig. \ref{fig:concurrence_plots1} and Fig. \ref{fig:concurrence_plots2}, it is easy to know that when $r_a = r_b = r$ the concurrences in physical accessible modes  $C(\rho_{A_IB_I})$  monotonically decreases with increasing Hawking temperature, but no sudden death occurs. However, the concurrences in physical inaccessible modes $C(\rho_{A_{II}B_{II}})$, $C(\rho_{A_IB_{II}})$  increase from zero as the Hawking temperature rises.

Meanwhile, Fig. \ref{fig:concurrence_plots2} shows that a higher monochromatic frequency corresponds to a lower physically accessible concurrence, while the physically inaccessible concurrence exhibits the opposite trend. This indicates that the energy generated by Hawking radiation has a disruptive effect on entanglement.

\section{IV. Concurrence under both noisy environments and Unruh effect }
In this section, the perturbation effects of three different decoherence channels on concurrence are further investigated.  In addition to the thermal noise from Hawking radiation, Alice and Bob are also subject to environmental noise. A separate noisy environment, whose properties are defined by specific couplings. The action of a noisy environment is described as
\begin{equation}\label{R19}
	\rho \rightarrow \rho^{evo}=\sum_k E_k \rho E_k^{\dagger},
\end{equation}
where $\rho(\rho^{evo})$ is the density matrix of a initial (final) state, $E_k$ ($E_k^{\dagger}$) is the single qubit Kraus (complex conjugate) operator of the noisy channel. Except the Unruh effect, let's move to discuss the interconnections among the formulas under various noisy environments, here we focus on phase damping channel, phase flip channel and bit flip channel as examples.

\subsection{A. In case of phase damping channel}
When we consider the case of the phase damping channel, the single qubit Kraus operators are given by,
\begin{equation}\label{R20}
\begin{aligned}
&E_1=\left(\begin{array}{cc}
1 & 0 \\
0 & \sqrt{1-k}
\end{array}\right),
&E_2=\left(\begin{array}{cc}
0 & 0 \\
0 & \sqrt{k}
\end{array}\right),\\
\end{aligned}
\end{equation}
where $k \in [0, 1]$ is a decay probability and in our study we assume that it depends only on time \cite{decay}.

Similarly, we can compute the evolved quantum states of the bipartite reduce states.
\begin{equation}
\begin{aligned}
\rho_{A_IB_I}=&\frac{1}{4}\Big(I_A\otimes I_B-\sin^2 r_a \sigma_3\otimes I_B-\sin^2 r_b I_A\otimes \sigma_3\\
&+\sqrt{1-k}p\cos r_a\cos r_b(\sigma_1\otimes \sigma_1+\sigma_2\otimes \sigma_2)\\
&+(\sin^2 r_a\sin^2 r_b-p\cos^2 r_a\cos^2 r_b)\sigma_3\otimes \sigma_3\Big),
\end{aligned}
\end{equation}

\begin{equation}
\begin{aligned}
\rho_{A_{II}B_{II}}=&\frac{1}{4}\Big(I_A\otimes I_B+\cos^2 r_a \sigma_3\otimes I_B+\cos^2 r_b I_A\otimes \sigma_3\\
&+\sqrt{1-k}p\sin r_a\sin r_b(\sigma_1\otimes \sigma_1+\sigma_2\otimes \sigma_2)\\
&+(\cos^2 r_a\cos^2 r_b-p\sin^2 r_a\sin^2 r_b)\sigma_3\otimes \sigma_3\Big),
\end{aligned}
\end{equation}

\begin{equation}
\begin{aligned}
\rho_{A_IB_{II}}=&\frac{1}{4}\Big(I_A\otimes I_B-\sin^2 r_a \sigma_3\otimes I_B+\cos^2 r_b I_A\otimes \sigma_3\\
&+\sqrt{1-k}p\cos r_a \sin r_b(\sigma_1\otimes \sigma_1-\sigma_2\otimes \sigma_2)\\
&-(\sin^2 r_a\cos^2 r_b-p\cos^2 r_a\sin^2 r_b)\sigma_3\otimes \sigma_3\Big),
\end{aligned}
\end{equation}

\begin{equation}
\begin{aligned}
\rho_{A_{II}B_{I}}=&\frac{1}{4}\Big(I_A\otimes I_B+\cos^2 r_a \sigma_3\otimes I_B-\sin^2 r_b I_A\otimes \sigma_3\\
&+\sqrt{1-k}p\sin r_a \cos r_b(\sigma_1\otimes \sigma_1-\sigma_2\otimes \sigma_2)\\
&-(\cos^2 r_a\sin^2 r_b-p\sin^2 r_a\cos^2 r_b)\sigma_3\otimes \sigma_3\Big).
\end{aligned}
\end{equation}

Employing Eq. (\ref{R11}), the concurrences of those states under phase damping noise are given by Eq. (\ref{pd}).
 \begin{figure*}
 \begin{subequations}\label{pd}
 \begin{align}
C_{pd}(\rho _{A_IB_I} )=\frac{1}{2}\left( 2p\sqrt{1-k}\cos r_a\cos r_b - \cos r_a\cos r_b\sqrt{(1-p)\left( 2\sin^2 r_a + 2\sin^2 r_b + (1-p)\cos^2 r_a\cos^2 r_b \right)} \right),\\
C_{pd}(\rho _{A_{II}B_{II}} )=\frac{1}{2}\left( 2p\sqrt{1-k}\sin r_a\sin r_b - \sin r_a\sin r_b\sqrt{(1-p)\left( 2\cos^2 r_a + 2\cos^2 r_b + (1-p)\sin^2 r_a\sin^2 r_b \right)} \right),\\
C_{pd}(\rho _{A_{I}B_{II}} )=\frac{1}{2}\left( 2p\sqrt{1-k}\cos r_a\sin r_b - \cos r_a\sin r_b\sqrt{(1-p)\left( 2\sin^2 r_a + 2\cos^2 r_b + (1-p)\cos^2 r_a\sin^2 r_b \right)} \right),\\
C_{pd}(\rho _{A_{II}B_{I}} )=\frac{1}{2}\left( 2p\sqrt{1-k}\sin r_a\cos r_b - \sin r_a\cos r_b\sqrt{(1-p)\left( 2\cos^2 r_a + 2\sin^2 r_b + (1-p)\sin^2 r_a\cos^2 r_b \right)} \right).
 \end{align}
 \end{subequations}
 \noindent \rule[-10pt]{18cm}{0.05em}
 \end{figure*}
In particular, the concurrences of the isotropic state for $p=1$ are given by
\begin{equation}\label{R22}
\begin{aligned}
&C_{pd}(\rho _{A_IB_I} )= \sqrt{1-k}\cos r_a\cos r_b,\\
&C_{pd}(\rho _{A_{II}B_{II}} )=\sqrt{1-k}\sin r_a\sin r_b,\\
&C_{pd}(\rho _{A_{I}B_{II}})=\sqrt{1-k}\cos r_a\sin r_b,\\
&C_{pd}(\rho _{A_{II}B_{I}} )=\sqrt{1-k}\sin r_a\cos r_b.
\end{aligned}
\end{equation}
From the analytical expression of concurrence given in Eqs. (\ref{R22}),  it can be seen that the concurrence is a function of the acceleration parameter  and the decay probability, all concurrences decrease monotonically as the decay probability \( k \) increases, and eventually vanish when \( k = 1 \).
Meanwhile, we can derive a trade-off relationship for the concurrence under phase damping channel noise,
\begin{equation}\label{R23}
\begin{aligned}
C_{pd}(\rho _{A_IB_I} )^2&+C_{pd}(\rho _{A_{II}B_{II}} )^2+C_{pd}(\rho _{A_{I}B_{II}} )^2\\
&+C_{pd}(\rho _{A_{II}B_{I}} )^2
=1-k.
\end{aligned}
\end{equation}
This complementarity captures the redistribution of quantum correlations among the observers with increasing acceleration, reflecting the conservation of quantum resources across different partitioning schemes.

\begin{figure*}[htbp]
	   	\centering
\vspace{-3.5cm}
		\includegraphics[width=0.4\textwidth]{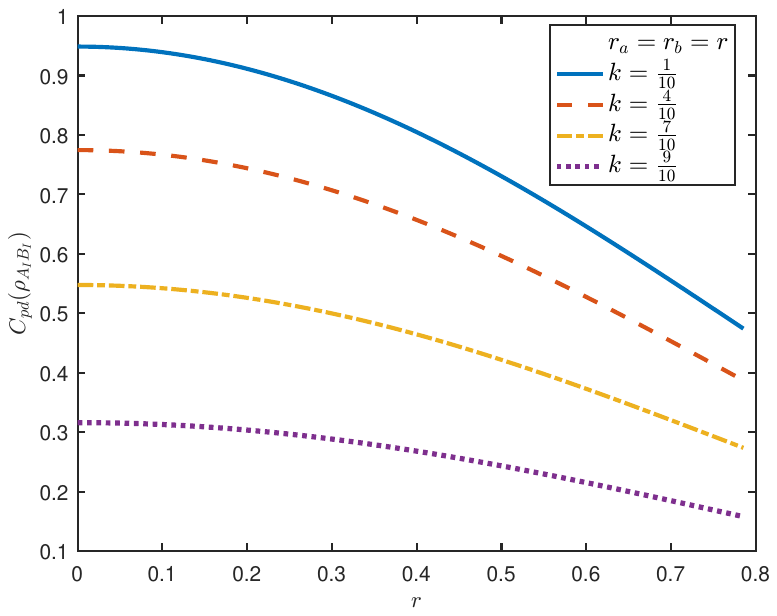}
\hspace{-2cm}
		\includegraphics[width=0.4\textwidth]{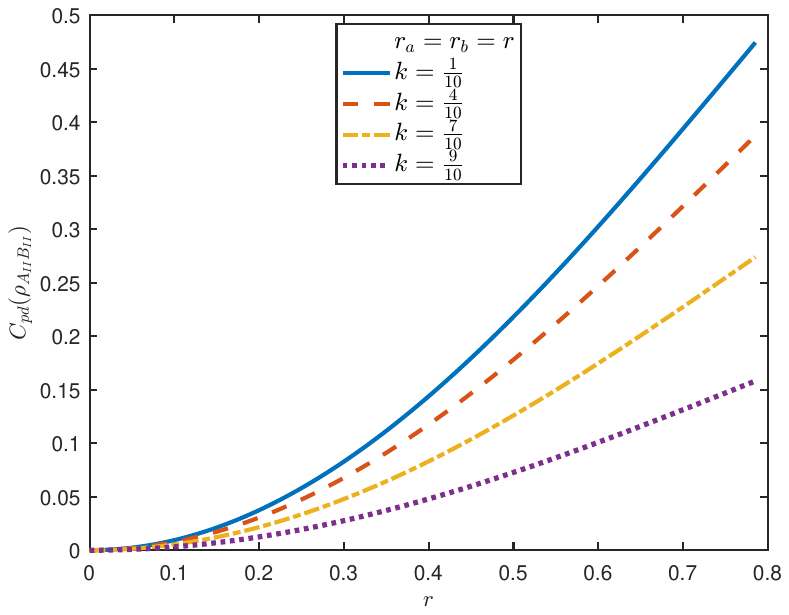}
\hspace{-2cm}
		\includegraphics[width=0.4\textwidth]{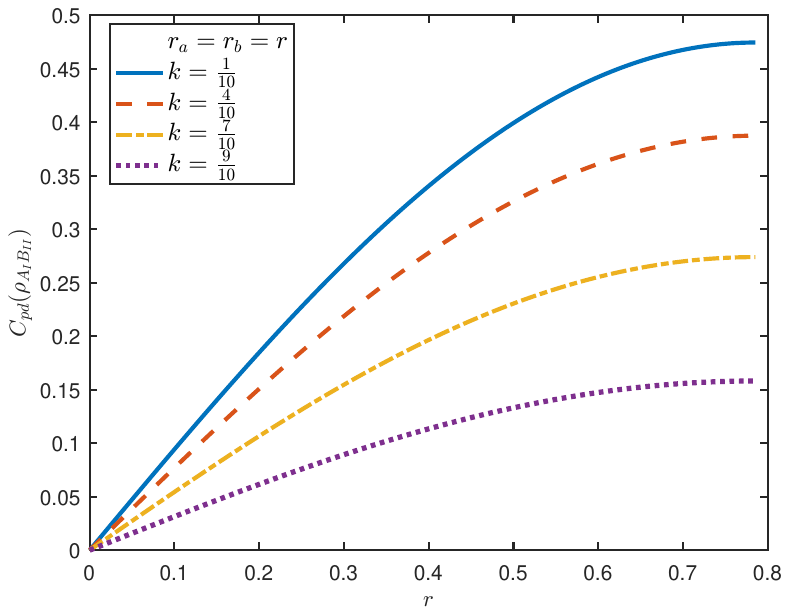}
\vspace{-3.5cm}
\caption{Plot $C_{pd}$ as a function of Hawking acceleration $r$ for different decay probability $k$ in case of $r_a = r_b = r$ and $p=1$ under phase damping channel.}
\label{fig:concurrence_plots3}
	   \end{figure*}

\begin{figure*}[htbp]
	   	\centering
\vspace{-3.5cm}
		\includegraphics[width=0.4\textwidth]{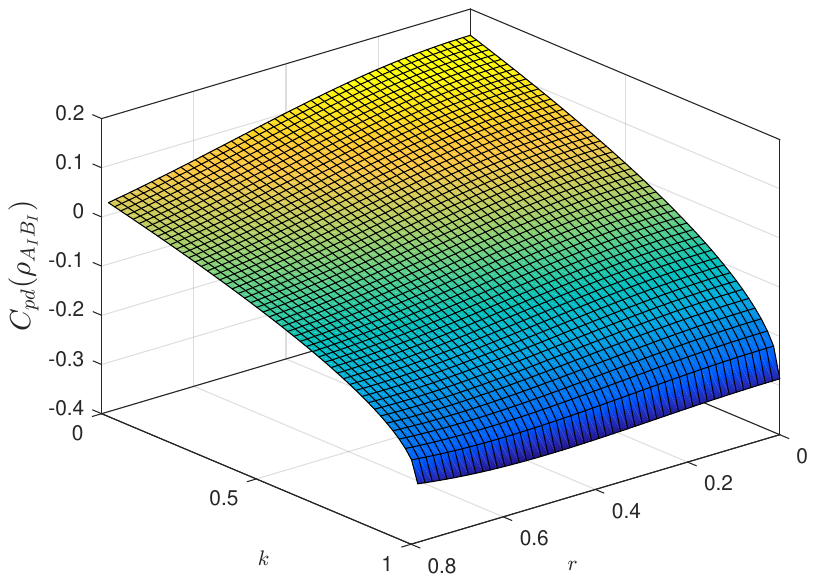}
\hspace{-2cm}
		\includegraphics[width=0.4\textwidth]{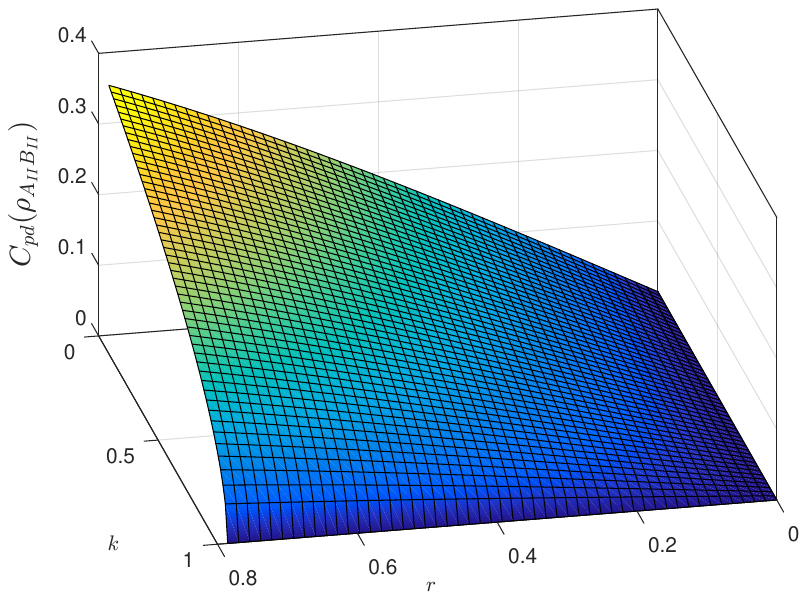}
\hspace{-2cm}
		\includegraphics[width=0.4\textwidth]{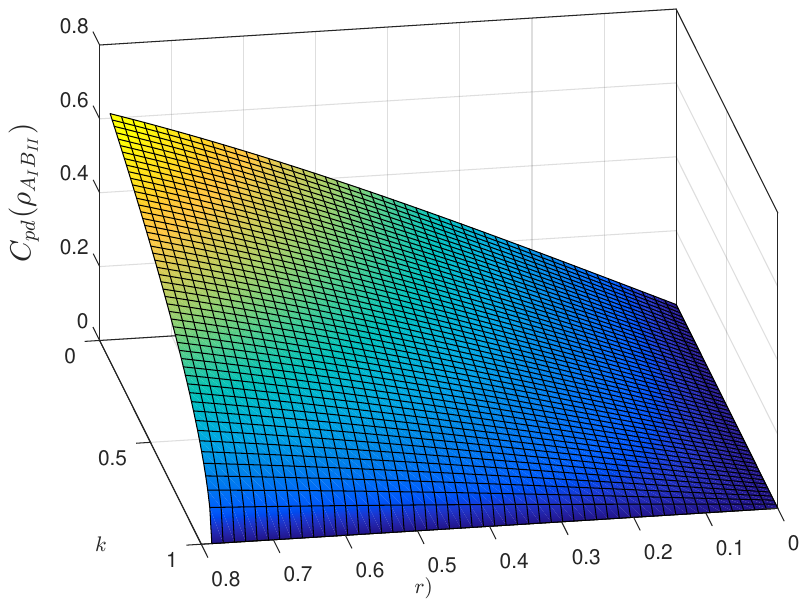}
\vspace{-3.5cm}
\caption{Plot $C_{pd}$ as functions of $r $ and $k$ when $r_a =\frac{\pi}{6}$, $r_b = r$, $p=1$ under phase damping noise.}
\label{fig:concurrence_plots5}
	   \end{figure*}

For a thorough analysis of the relationship between concurrence and the acceleration, state parameters, and channel noise parameters, we have constructed plots of the concurrence function (\ref{R22}) in Fig. \ref{fig:concurrence_plots3} and Fig. \ref{fig:concurrence_plots5}. As shown in Fig. \ref{fig:concurrence_plots3}, the concurrence in the physically accessible mode \( A_I B_I \) decreases as the Hawking acceleration increases, yet it does not vanish even at maximal acceleration \( r=\frac{\pi}{4} \). In contrast, the concurrences in the physically inaccessible modes---\( A_{II}B_{II} \), \( A_{I}B_{II} \), and \( A_{II}B_{I} \), each involving both parties \( A \) and \( B \)---increase with higher Hawking acceleration.

 In addition, Fig. \ref{fig:concurrence_plots5} shows another interesting phenomenon: both the physically accessible and inaccessible concurrencies decrease monotonically with the increase of channel damping. The sudden death occurs when k=1, which indicates that the influence of channel noise is greater than that of Hawking acceleration.

\subsection{B. In case of phase flip channel}
We now turn to the evolution of concurrence for a quantum state passing through a phase flip channel.
In the case of the phase flip channel, the single qubit Kraus operators are given by
\begin{equation}\label{R24}
\begin{aligned}
&E_1=\left(\begin{array}{cc}
\sqrt{1-k} & 0 \\
0 & \sqrt{1-k}
\end{array}\right),
&E_2=\left(\begin{array}{cc}
\sqrt{k} & 0 \\
0 & -\sqrt{k}
\end{array}\right).
\end{aligned}
\end{equation}

Using the same method, we can calculate the reduced states of the quantum states after passing through the phase flip channel.
\begin{equation}
\begin{aligned}
\rho_{A_IB_I}=&\frac{1}{4}\Big(I_A\otimes I_B-\sin^2 r_a \sigma_3\otimes I_B-\sin^2 r_b I_A\otimes \sigma_3\\
&+(1-2k)p\cos r_a\cos r_b(\sigma_1\otimes \sigma_1+\sigma_2\otimes \sigma_2)\\
&+(\sin^2 r_a\sin^2 r_b-p\cos^2 r_a\cos^2 r_b)\sigma_3\otimes \sigma_3\Big),
\end{aligned}
\end{equation}

\begin{equation}
\begin{aligned}
\rho_{A_{II}B_{II}}=&\frac{1}{4}\Big(I_A\otimes I_B+\cos^2 r_a \sigma_3\otimes I_B+\cos^2 r_b I_A\otimes \sigma_3\\
&+(1-2k)p\sin r_a\sin r_b(\sigma_1\otimes \sigma_1+\sigma_2\otimes \sigma_2)\\
&+(\cos^2 r_a\cos^2 r_b-p\sin^2 r_a\sin^2 r_b)\sigma_3\otimes \sigma_3\Big),
\end{aligned}
\end{equation}

\begin{equation}
\begin{aligned}
\rho_{A_IB_{II}}=&\frac{1}{4}\Big(I_A\otimes I_B-\sin^2 r_a \sigma_3\otimes I_B+\cos^2 r_b I_A\otimes \sigma_3\\
&+(1-2k)p\cos r_a \sin r_b(\sigma_1\otimes \sigma_1-\sigma_2\otimes \sigma_2)\\
&-(\sin^2 r_a\cos^2 r_b-p\cos^2 r_a\sin^2 r_b)\sigma_3\otimes \sigma_3\Big),
\end{aligned}
\end{equation}

\begin{equation}
\begin{aligned}
\rho_{A_{II}B_{I}}=&\frac{1}{4}\Big(I_A\otimes I_B+\cos^2 r_a \sigma_3\otimes I_B-\sin^2 r_b I_A\otimes \sigma_3\\
&+(1-2k)p\sin r_a \cos r_b(\sigma_1\otimes \sigma_1-\sigma_2\otimes \sigma_2)\\
&-(\cos^2 r_a\sin^2 r_b-p\sin^2 r_a\cos^2 r_b)\sigma_3\otimes \sigma_3\Big).
\end{aligned}
\end{equation}

Also, we calculate the concurrences of $\rho$ after influence by the phase flip channel,
in case of $k\in [0, \frac{1}{2}]$, we obtain the concurrences in Eq. (\ref{pf})
 \begin{figure*}
 \begin{subequations}\label{pf}
 \begin{align}
C_{pf}(\rho _{A_IB_I} )=\frac{1}{2}\left( 2p(1-2k)\cos r_a\cos r_b - \cos r_a\cos r_b\sqrt{(1-p)\left( 2\sin^2 r_a + 2\sin^2 r_b + (1-p)\cos^2 r_a\cos^2 r_b \right)} \right),\\
C_{pf}(\rho _{A_{II}B_{II}} )=\frac{1}{2}\left( 2p(1-2k)\sin r_a\sin r_b - \sin r_a\sin r_b\sqrt{(1-p)\left( 2\cos^2 r_a + 2\cos^2 r_b + (1-p)\sin^2 r_a\sin^2 r_b \right)} \right),\\
C_{pf}(\rho _{A_{I}B_{II}} )=\frac{1}{2}\left( 2p(1-2k)\cos r_a\sin r_b - \cos r_a\sin r_b\sqrt{(1-p)\left( 2\sin^2 r_a + 2\cos^2 r_b + (1-p)\cos^2 r_a\sin^2 r_b \right)} \right),\\
C_{pf}(\rho _{A_{II}B_{I}} )=\frac{1}{2}\left( 2p(1-2k)\sin r_a\cos r_b - \sin r_a\cos r_b\sqrt{(1-p)\left( 2\cos^2 r_a + 2\sin^2 r_b + (1-p)\sin^2 r_a\cos^2 r_b \right)} \right).
 \end{align}
 \end{subequations}
 \noindent \rule[-10pt]{18cm}{0.05em}
 \end{figure*}
In particular, the concurrence of the isotropic state for p=1 is given by
\begin{equation}\label{R26}
\begin{aligned}
&C_{pf}(\rho _{A_IB_I} )= (1-2k)\cos r_a\cos r_b,\\
&C_{pf}(\rho _{A_{II}B_{II}} )=(1-2k)\sin r_a\sin r_b,\\
&C_{pf}(\rho _{A_{I}B_{II}} )=(1-2k)\cos r_a\sin r_b,\\
&C_{pf}(\rho _{A_{II}B_{I}} )=(1-2k)\sin r_a\cos r_b.
\end{aligned}
\end{equation}
and all the concurrences reduce to zero for $k\in [0,\frac{1}{2}]$.

These expressions explicitly provide the mathematical formulas for the concurrence of each reduced state under the influence of phase flip channel noise. The complementarity relation among the four concurrences is directly captured as expressed in the following formula,
\begin{equation}\label{R27}
\begin{aligned}
C_{pf}(\rho _{A_IB_I} )^2&+C_{pf}(\rho _{A_{II}B_{II}} )^2+C_{pf}(\rho _{A_{I}B_{II}} )^2\\&+C_{pf}(\rho _{A_{II}B_{I}} )^2=(1-2k)^2.
\end{aligned}
\end{equation}

\begin{figure*}[htbp]
	\centering
\vspace{-3.5cm}
		\includegraphics[width=0.4\textwidth]{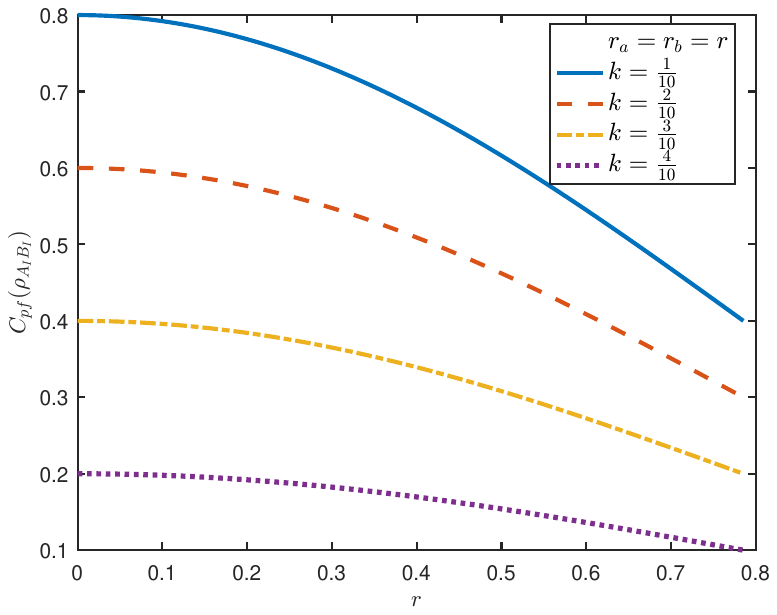}
\hspace{-2cm}
		\includegraphics[width=0.4\textwidth]{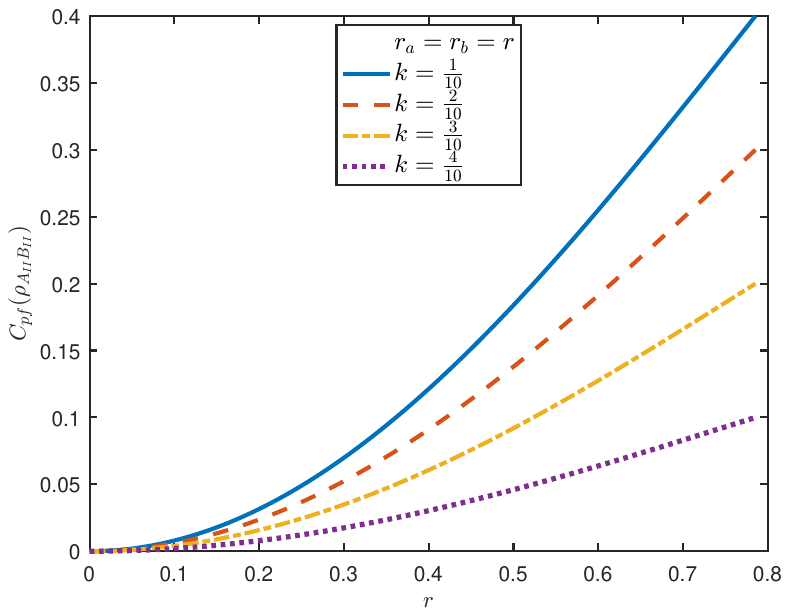}
\hspace{-2cm}
		\includegraphics[width=0.4\textwidth]{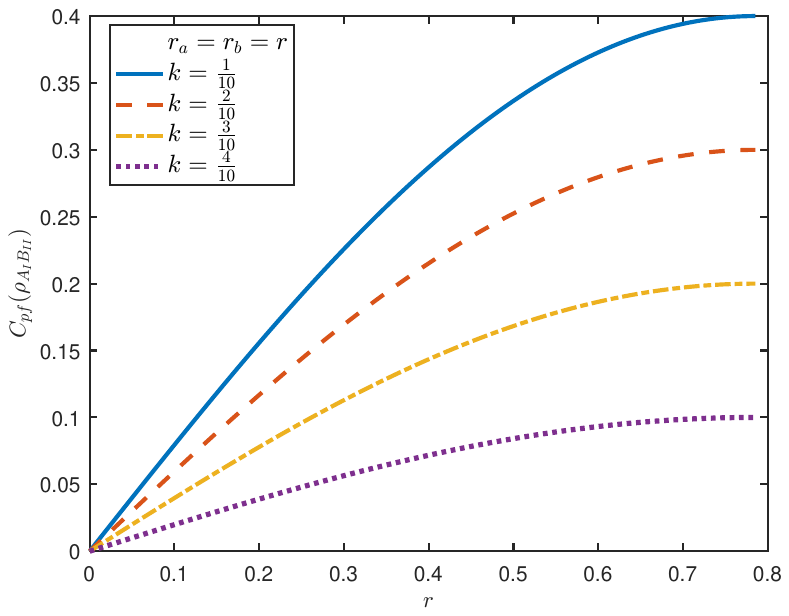}
\vspace{-3.5cm}
\caption{Plot $C_{pf}$ as a function of Hawking acceleration $r$ for different decay probability $k$ in case of $r_a = r_b = r$ and $p=1$  under phase flip channel.}
\label{fig:concurrence_plots7}
\end{figure*}

\begin{figure*}[htbp]
	\centering
\vspace{-3.5cm}
		\includegraphics[width=0.4\textwidth]{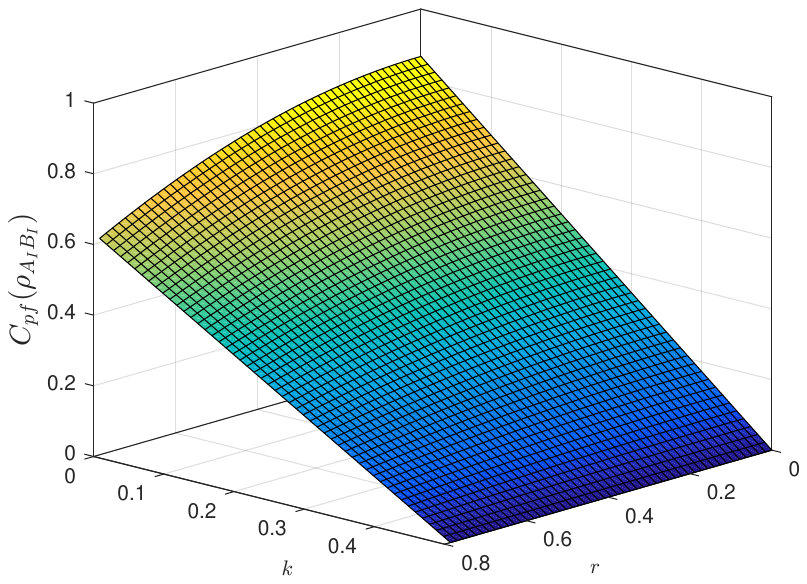}
\hspace{-2cm}
		\includegraphics[width=0.4\textwidth]{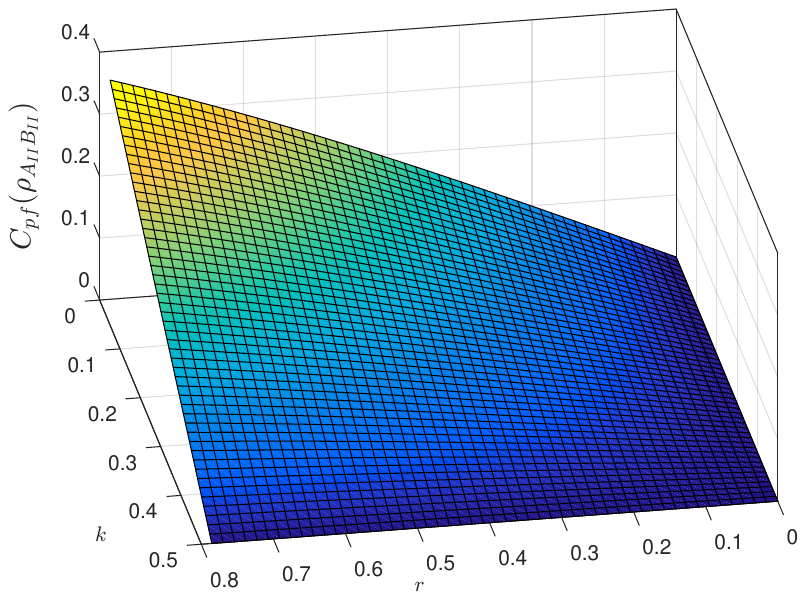}
\hspace{-2cm}
		\includegraphics[width=0.4\textwidth]{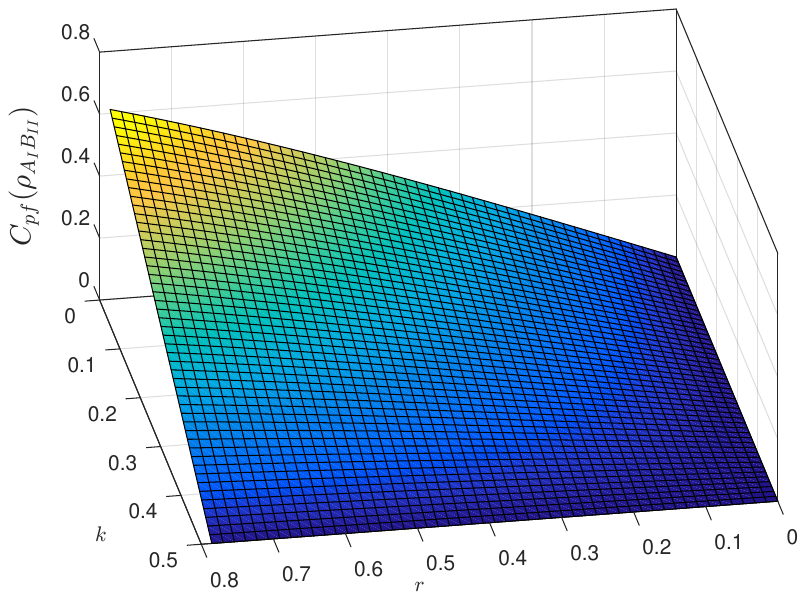}
\vspace{-3.5cm}
\caption{Plot $C_{pf}$ as functions of $r $ and $k$ when $r_a =\frac{\pi}{6}$, $r_b = r$,$p=1$ under phase flip channel.}
\label{fig:concurrence_plots9}
\end{figure*}

In a similar manner, we plot the concurrence function (\ref{R26}) for the quantum state after its passage through the phase flip channel to further investigate its relationship with the Hawking acceleration, state parameters, and channel noise parameters in Fig. \ref{fig:concurrence_plots7} and Fig. \ref{fig:concurrence_plots9}. As shown in Fig. \ref{fig:concurrence_plots7} , with increasing acceleration, the physically inaccessible concurrence under phase flip channel gradually increases from zero. The larger the channel damping coefficient, the slower the growth rate of concurrence. In contrast, the physically accessible concurrence exhibits the opposite trend: it starts to decrease from a fixed value, and a larger channel damping coefficient results in a slower decline in concurrence.

Furthermore, Fig. \ref{fig:concurrence_plots9} illustrates that as the channel damping rate increases, both the physically accessible and inaccessible concurrence diminish. When $k\geq \frac{1}{2}$,
the concurrence drops to zero, at which point the phenomenon of sudden death occurs.

\subsection{C. In case of bit flip channel}
Next, we examine the evolution of concurrence through a bit flip channel. In case of the bit flip channel, the single qubit Kraus operators for it are given by,
\begin{equation}\label{R28}
\begin{aligned}
&E_1=\left(\begin{array}{cc}
\sqrt{1-k} & 0 \\
0 & \sqrt{1-k}
\end{array}\right),
&E_2=\left(\begin{array}{cc}
0 & \sqrt{k} \\
\sqrt{k} & 0
\end{array}\right).\\
\end{aligned}
\end{equation}

Using a similar approach, we can calculate the respective reduced states of the quantum states after passing through the bit flip channel.
\begin{equation}
\begin{aligned}
\rho&_{A_IB_I}=\frac{1}{4}\Big(I_A\otimes I_B-\sin^2 r_a \sigma_3\otimes I_B-(1-2k)\sin^2 r_b\\& I_A\otimes \sigma_3
+p\cos r_a\cos r_b(\sigma_1\otimes \sigma_1 +(1-2k)
\sigma_2\otimes \sigma_2)
+\\&(1-2k)(\sin^2 r_a\sin^2 r_b-p\cos^2 r_a\cos^2 r_b)\sigma_3\otimes \sigma_3\Big),
\end{aligned}
\end{equation}

\begin{equation}
\begin{aligned}
\rho&_{A_{II}B_{II}}=\frac{1}{4}\Big(I_A\otimes I_B+\cos^2 r_a \sigma_3\otimes I_B+(1-2k)\cos^2 r_b \\&I_A\otimes \sigma_3+p\sin r_a\sin r_b(\sigma_1\otimes \sigma_1+(1-2k)\sigma_2\otimes \sigma_2)+\\&(1-2k)(\cos^2 r_a\cos^2 r_b-p\sin^2 r_a\sin^2 r_b)\sigma_3\otimes \sigma_3\Big),
\end{aligned}
\end{equation}

\begin{equation}
\begin{aligned}
\rho&_{A_IB_{II}}=\frac{1}{4}\Big(I_A\otimes I_B-\sin^2 r_a \sigma_3\otimes I_B+(1-2k)\cos^2 r_b \\&I_A\otimes \sigma_3+p\cos r_a \sin r_b(\sigma_1\otimes \sigma_1-(1-2k)p\sigma_2\otimes \sigma_2
-\\&(1-2k)(\sin^2 r_a\cos^2 r_b-p\cos^2 r_a\sin^2 r_b)\sigma_3\otimes \sigma_3\Big),
\end{aligned}
\end{equation}

\begin{equation}
\begin{aligned}
\rho&_{A_{II}B_{I}}=\frac{1}{4}\Big(I_A\otimes I_B+\cos^2 r_a \sigma_3\otimes I_B-(1-2k)\sin^2 r_b \\&I_A\otimes \sigma_3+p\sin r_a \cos r_b(\sigma_1\otimes \sigma_1-(1-2k)\sigma_2\otimes \sigma_2)
-\\&(1-2k)(\cos^2 r_a\sin^2 r_b-p\sin^2 r_a\cos^2 r_b)\sigma_3\otimes \sigma_3\Big).
\end{aligned}
\end{equation}

Compared to phase damping and phase flip channels, the bit flip channel has a more complex effect on the concurrence. This is primarily due to the strong dependence of the four eigenvalues of the reduced state on the noise strength parameter $k$, which leads to complicated changes in their ordering and necessitates separate discussions for different parameter regimes. This, in turn, significantly increases the complexity of calculating the concurrence of the reduced states. Accordingly, we provide the expression for concurrence in the appendix and focus on the numerical analysis in the main text.

\begin{figure*}[htbp]
	\centering
\vspace{-3.5cm}
		\includegraphics[width=0.4\textwidth]{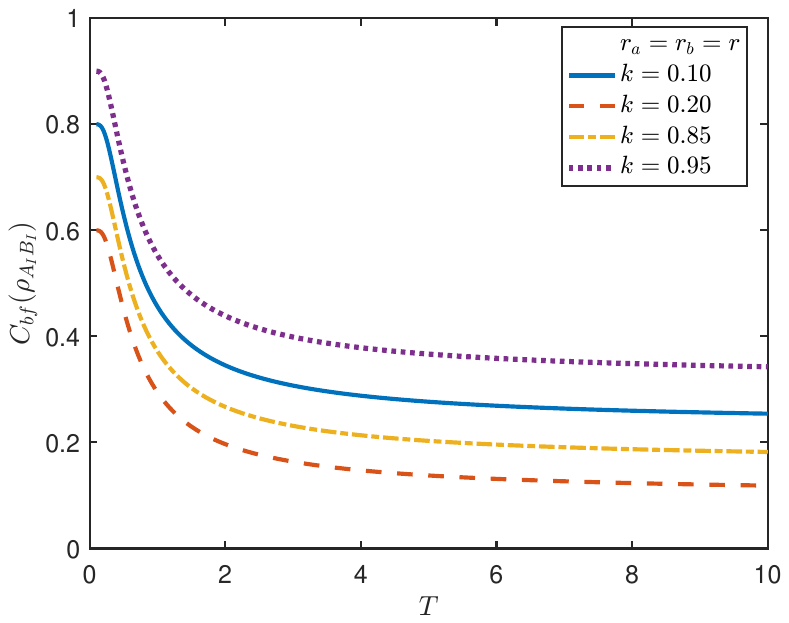}
\hspace{-2cm}
		\includegraphics[width=0.4\textwidth]{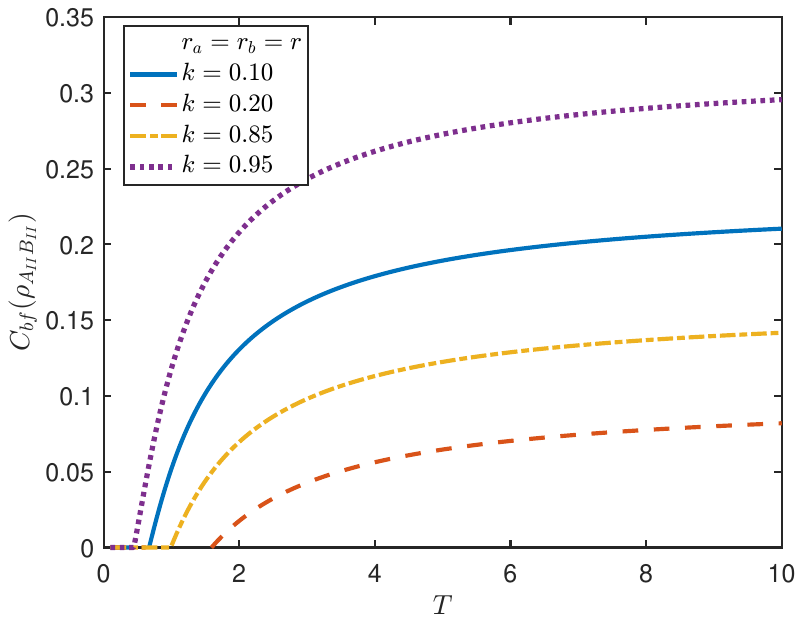}
\hspace{-2cm}
		\includegraphics[width=0.4\textwidth]{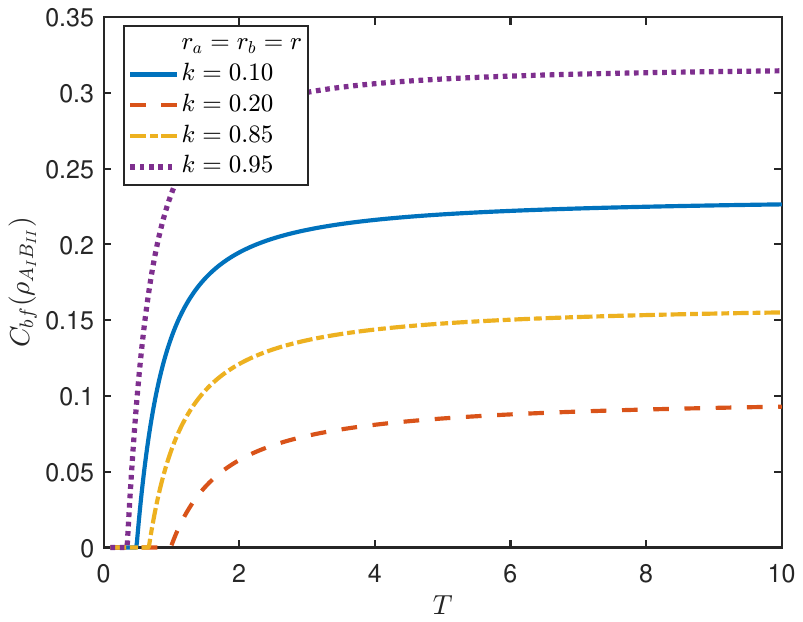}
\vspace{-3.5cm}
\caption{Concurrence $C_{bf}$ versus Hawking temperature under the bit flip channel. These panels show how the concurrence evolves with the Hawking temperature under the influence of a bit flip channel
when Alice and Bob fall into the black hole with the same acceleration and $\omega=1$,$p=1$.}
\label{fig:concurrence_plots14}
\end{figure*}

As shown in Fig. \ref{fig:concurrence_plots14}, the physically accessible concurrence undergoes a monotonic decrease with Hawking temperature under the bit flip channel, characterized by a rapid initial decline followed by a slower decay, yet without exhibiting sudden death. Conversely, the physically inaccessible concurrence shows the precisely opposite behavior.
\begin{figure*}[htbp]
	\centering
\vspace{-3.5cm}
		\includegraphics[width=0.4\textwidth]{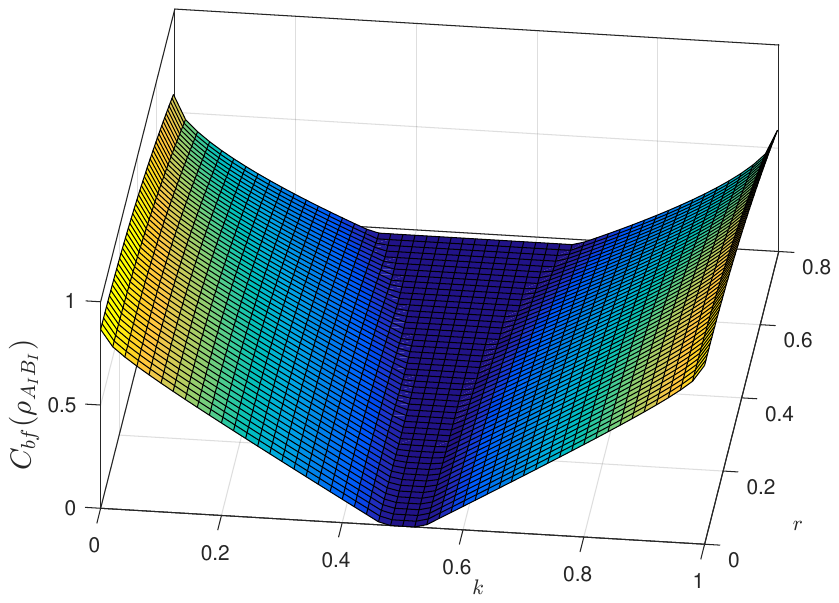}
\hspace{-2cm}
		\includegraphics[width=0.4\textwidth]{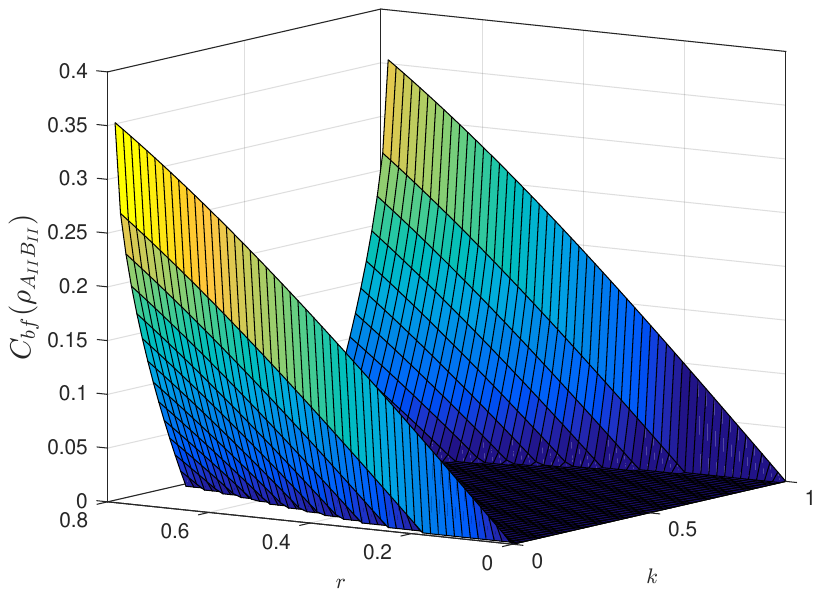}
\hspace{-2cm}
		\includegraphics[width=0.4\textwidth]{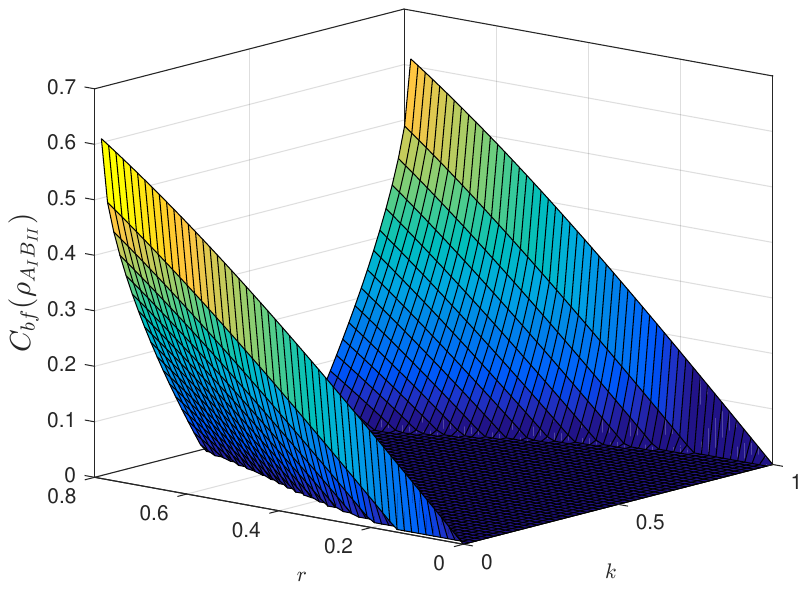}
\vspace{-3.5cm}
\caption{Plot the concurrence $C_{bf}$ as a function of both the Hawking acceleration parameter $r$ and the noise strength $k$. It shows how the concurrences evolve for Alice and Bob undergoing different accelerations near a black hole (i.e. $r_a =\frac{\pi}{6}$, $r_b = r$,$p=1$) and subject to a bit-flip channel.}
\label{fig:concurrence_plots15}
\end{figure*}
Fig. \ref{fig:concurrence_plots15} shows that under a bit flip channel, the concurrence for the physically accessible and inaccessible  exhibits a symmetric dependence on the noise strength $k$. This behavior differs from the cases of phase damping and phase flip channels. Additionally, Fig. \ref{fig:concurrence_plots15} reveals symmetric characteristics in concurrency during channel noise evolution.

\section{V. Conclusions}
It is well established that quantum entanglement captures the essence of quantum mechanics and plays a crucial role in quantum information processing. As a quantum correlation, it characterizes the fundamental nature of quantum states. In this work, we employ concurrence as an entanglement measure and select a two-qubit mixed state as the initial state to investigate its dynamics under Hawking radiation and three types of channel noise.
Based on the derived analytical expression for concurrence, we establish some trade-off relations between the physically accessible and inaccessible concurrence. This constraint delineates a complementary relationship: the total concurrence is conserved, such that a decrease in one part is invariably accompanied by an increase in the other.

Furthermore, our findings indicate that as the Hawking temperature rises, the physically accessible concurrence decreases monotonically, while the inaccessible part increases monotonically. Thus, our study of entanglement evolution based on mixed states in black holes coincides with previous results derived from pure states \cite{qiang2018, wsm2024, kim}, namely, Hawking radiation exerts a decoherence effect on entanglement. However, the robustness of entanglement for mixed states against channel noise differs from that of pure states, with notable divergence in performance within bit flip channels.
Sudden death occurs and the concurrence exhibits a distinct symmetry with respect to the channel's noise parameter. This indicates that the channel noise has a more pronounced impact on entanglement than Hawking radiation.

\bigskip

\noindent{\bf Acknowledgments}
This work is supported by
the Natural Science Foundation of Hainan Province under Grant No. 125RC744; the China Scholarship Council (CSC); the specific research fund of the Innovation Platform for Academicians of Hainan Province.

\begin{widetext}
\section{Appendix}
For the initial isotropic  state $\rho = \frac{1-p}{4} I \otimes I + p |\psi^+\rangle\langle\psi^+|$, with two observers Alice and Bob located in the event horizon region and simultaneously affected to the channel noise, the four reduced density matrices obtained after performing the partial trace over two subsystems have different mathematical forms, furthermore, we can obtain the their concurrences by using Eq. (\ref{R11}).


\begin{equation}
C_{bf}(\rho _{A_IB_I} ) =
\begin{cases}
(1-k)p\cos r_a\cos r_b \\
- \frac{1}{2}\sqrt{\cos^2r_a[2k+t(1-p)\cos^2r_b][(1+\sin^2r_a)(1+t\sin^2r_b)-tp\cos^2r_a\cos^2r_b]}, & k\in[0, \omega_1]; \\
0, & k\in(\omega_1, \omega_2);\\
kp\cos r_a\cos r_b  \\
-\frac{1}{2}\sqrt{\cos^2r_a[2(1-k)-t(1-p)\cos^2r_b][(1+\sin^2r_a)(1-t\sin^2r_b)+tp\cos^2r_a\cos^2r_b]}, & k\in[\omega_2, 1].
\end{cases}
\end{equation}

\begin{equation}
C_{bf}(\rho _{A_{II}B_{II}} )=
\begin{cases}
(1-k)p\sin r_a\sin r_b \\
- \frac{1}{2}\sqrt{\sin^2r_a[2k+t(1-p)\sin^2r_b][(1+\cos^2r_a)(1+t\cos^2r_b)-tp\sin^2r_a\sin^2r_b]}, & k\in[0, \omega_3]; \\
0, & k\in(\omega_3, \omega_4);\\
kp\sin r_a\sin r_b  \\
-\frac{1}{2}\sqrt{\sin^2r_a[2(1-k)-t(1-p)\sin^2r_b][(1+\cos^2r_a)(1-t\cos^2r_b)+tp\sin^2r_a\sin^2r_b]}, & k\in[\omega_4, 1].
\end{cases}
\end{equation}

\begin{equation}
C_{bf}(\rho _{A_{I}B_{II}} )=
\begin{cases}
(1-k)p\cos r_a\sin r_b \\
- \frac{1}{2}\sqrt{\cos^2r_a[2k+t(1-p)\sin^2r_b][(1+\sin^2r_a)(1+t\cos^2r_b)-tp\cos^2r_a\sin^2r_b]}, & k\in[0, \omega_5]; \\
0, & k\in(\omega_5, \omega_6);\\
kp\cos r_a\sin r_b  \\
-\frac{1}{2}\sqrt{\cos^2r_a[2(1-k)-t(1-p)\sin^2r_b][(1+\sin^2r_a)(1-t\cos^2r_b)+tp\cos^2r_a\sin^2r_b]}, & k\in[\omega_6, 1].
\end{cases}
\end{equation}

\begin{equation}
C_{bf}(\rho _{A_{II}B_{I}} )=
\begin{cases}
(1-k)p\sin r_a\cos r_b \\
- \frac{1}{2}\sqrt{\sin^2r_a[2k+t(1-p)\cos^2r_b][(1+\cos^2r_a)(1+t\sin^2r_b)-tp\sin^2r_a\cos^2r_b]}, & k\in[0, \omega_7]; \\
0, & k\in(\omega_7, \omega_8);\\
kp\sin r_a\cos r_b  \\
-\frac{1}{2}\sqrt{\sin^2r_a[2(1-k)-t(1-p)\cos^2r_b][(1+\cos^2r_a)(1-t\sin^2r_b)+tp\sin^2r_a\cos^2r_b]}, & k\in[\omega_8, 1].
\end{cases}
\end{equation}

where
\begin{equation}
\begin{aligned}
\omega_1 =\frac{-a_{12}-\sqrt{a_{12}^2-4a_{11}a_{13}}}{2a_{11}}, & \omega_2 =\frac{-a_{22}+\sqrt{a_{22}^2-4a_{21}a_{23}}}{2a_{21}}, & \omega_3 =\frac{-b_{12}-\sqrt{b_{12}^2-4b_{11}b_{13}}}{2b_{11}},\\
\omega_4 =\frac{-b_{22}+\sqrt{b_{22}^2-4b_{21}b_{23}}}{2b_{21}}, &\omega_5 =\frac{-c_{12}-\sqrt{c_{12}^2-4c_{11}c_{13}}}{2c_{11}}, &\omega_6 =\frac{-c_{22}+\sqrt{c_{22}^2-4c_{21}c_{23}}}{2c_{21}},\\
\omega_7 =\frac{-d_{12}-\sqrt{d_{12}^2-4d_{11}d_{13}}}{2d_{11}}, &\omega_8 =\frac{-d_{22}+\sqrt{d_{22}^2-4d_{21}d_{23}}}{2d_{21}}, &  t=1-2k.\\
\end{aligned}
\end{equation}


where the parameters are given by
\begin{equation}
\begin{aligned}
a_{11}&=4p^2\cos^2r_b+4(1-p)\cos^2r_b[(1+\sin^2r_a)\sin^2r_b-p\cos^2r_a\sin^2r_b],\\
a_{12}&=-8p^2\cos^2r_b-2(1-p)\cos^2r_b[(1+\sin^2r_a)\sin^2r_b-p\cos^2r_a\sin^2r_b]-2[(1+\sin^2r_a)(1+\sin^2r_b)-p\cos^2r_a\sin^2r_b],\\
a_{13}&=4p^2\cos^2r_b-(1-p)\cos^2r_b[(1+\sin^2r_a)(1+\sin^2r_b)-p\cos^2r_a\sin^2r_b].\\
a_{21}&=4p^2\cos^2r_b-4(1-p)\cos^2r_b[(1+\sin^2r_a)\sin^2r_b-p\cos^2r_a\sin^2r_b],\\
a_{22}&=-2(1-p)\cos^2r_b[(1+\sin^2r_a)\sin^2r_b-p\cos^2r_a\sin^2r_b]-2[(1+\sin^2r_a)(1-\sin^2r_b)+p\cos^2r_a\sin^2r_b],\\
\end{aligned}
\end{equation}

\begin{equation}
\begin{aligned}
b_{11}&=4p^2\sin^2r_b+4(1-p)\sin^2r_b[(1+\cos^2r_a)\cos^2r_b-p\sin^2r_a\cos^2r_b],\\
b_{12}&=-8p^2\sin^2r_b-2(1-p)\sin^2r_b[(1+\cos^2r_a)\cos^2r_b-p\sin^2r_a\cos^2r_b]-2[(1+\cos^2r_a)(1+\cos^2r_b)-p\sin^2r_a\cos^2r_b],\\
b_{13}&=4p^2\sin^2r_b-(1-p)\sin^2r_b[(1+\cos^2r_a)(1+\cos^2r_b)-p\sin^2r_a\cos^2r_b].\\
b_{21}&=4p^2\sin^2r_b-4(1-p)\sin^2r_b[(1+\cos^2r_a)\cos^2r_b-p\sin^2r_a\cos^2r_b],\\
b_{22}&=-2(1-p)\sin^2r_b[(1+\cos^2r_a)\cos^2r_b-p\sin^2r_a\cos^2r_b]-2[(1+\cos^2r_a)(1-\cos^2r_b)+p\sin^2r_a\cos^2r_b],\\
b_{23}&=-2(1+\cos^2r_a)+(1-p)\sin^2r_b[(1+\cos^2r_a)(1-\cos^2r_b)+p\sin^2r_a\cos^2r_b].\\
\end{aligned}
\end{equation}

\begin{equation}
\begin{aligned}
c_{11}&=4p^2\sin^2r_b+4(1-p)\sin^2r_b[(1+\sin^2r_a)\cos^2r_b-p\cos^2r_a\sin^2r_b],\\
c_{12}&=-8p^2\sin^2r_b-2(1-p)\sin^2r_b[(1+\sin^2r_a)\cos^2r_b-p\cos^2r_a\sin^2r_b]-2[(1+\sin^2r_a)(1+\cos^2r_b)-p\cos^2r_a\sin^2r_b],\\
c_{13}&=4p^2\sin^2r_b-(1-p)\sin^2r_b[(1+\sin^2r_a)(1+\cos^2r_b)-p\cos^2r_a\sin^2r_b].\\
c_{21}&=4p^2\sin^2r_b-4(1-p)\sin^2r_b[(1+\sin^2r_a)\cos^2r_b-p\cos^2r_a\sin^2r_b],\\
c_{22}&=-2(1-p)\sin^2r_b[(1+\sin^2r_a)\cos^2r_b-p\cos^2r_a\sin^2r_b]-2[(1+\sin^2r_a)(1-\cos^2r_b)+p\cos^2r_a\sin^2r_b],\\
c_{23}&=-2(1+\sin^2r_a)+(1-p)\sin^2r_b[(1+\sin^2r_a)(1-\cos^2r_b)+p\cos^2r_a\sin^2r_b].\\
\end{aligned}
\end{equation}


\begin{equation}
\begin{aligned}
d_{11}&=4p^2\cos^2r_b+4(1-p)\cos^2r_b[(1+\cos^2r_a)\sin^2r_b-p\sin^2r_a\cos^2r_b],\\
d_{12}&=-8p^2\cos^2r_b-2(1-p)\cos^2r_b[(1+\cos^2r_a)\sin^2r_b-p\sin^2r_a\cos^2r_b]-2[(1+\cos^2r_a)(1+\sin^2r_b)-p\sin^2r_a\cos^2r_b],\\
d_{13}&=4p^2\cos^2r_b-(1-p)\cos^2r_b[(1+\cos^2r_a)(1+\sin^2r_b)-p\sin^2r_a\cos^2r_b].\\
d_{21}&=4p^2\cos^2r_b-4(1-p)\cos^2r_b[(1+\cos^2r_a)\sin^2r_b-p\sin^2r_a\cos^2r_b],\\
d_{22}&=-2(1-p)\cos^2r_b[(1+\cos^2r_a)\sin^2r_b-p\sin^2r_a\cos^2r_b]-2[(1+\cos^2r_a)(1-\sin^2r_b)+p\sin^2r_a\cos^2r_b],\\
d_{23}&=-2(1+\cos^2r_a)+(1-p)\cos^2r_b[(1+\cos^2r_a)(1-\sin^2r_b)+p\sin^2r_a\cos^2r_b].\\
\end{aligned}
\end{equation}

When $p=1$, the isotropic state degenerates into a maximally entangled pure state, at which point the concurrences of each of its reduced states become
\begin{equation}\label{R30}
C_{bf}(\rho _{A_IB_I}) =
\begin{cases}
(1-k)\cos r_a\cos r_b- \sqrt{k\cos^2r_a[k+(1-k)\sin^2r_a+(1-2k)\sin^2r_b]}, & k\in[0, \omega_1)\\
k\cos r_a\cos r_b- \sqrt{(1-k)\cos^2r_a[1-k+k\sin^2r_a-(1-2k)\sin^2r_b]}, & k\in(\omega_2, 1]\\
0, & k\in(\omega_1, \omega_2)
\end{cases}
\end{equation}

\begin{equation}\label{R31}
C_{bf}(\rho _{A_{II}B_{II}} )=
\begin{cases}
(1-k)\sin r_a\sin r_b - \sqrt{k\sin^2r_a[k+(1-k)\cos^2r_a+(1-2k)\cos^2r_b]}, & k\in[0, \omega_3) \\
k\sin r_a\sin r_b - \sqrt{(1-k)\sin^2r_a[1-k+k\cos^2r_a-(1-2k)\cos^2r_b]}, & k\in(\omega_4, 1] \\
0, & k\in(\omega_3, \omega_4)
\end{cases}
\end{equation}

\begin{equation}\label{R32}
C_{bf}(\rho _{A_{I}B_{II}} )=
\begin{cases}
(1-k)\cos r_a\sin r_b - \sqrt{k\cos^2r_a[1-k+(1-k)\sin^2r_a-(1-2k)\sin^2r_b]}, & k\in[0, \omega_5) \\
k\cos r_a\sin r_b - \sqrt{(1-k)\cos^2r_a[k+k\sin^2r_a+(1-2k)\sin^2r_b]}, & k\in(\omega_6, 1] \\
0, & k\in(\omega_5, \omega_6)
\end{cases}
\end{equation}

\begin{equation}\label{R33}
C_{bf}(\rho _{A_{II}B_{I}})=
\begin{cases}
(1-k)\sin r_a\cos r_b - \sqrt{k\sin^2r_a[1-k+(1-k)\cos^2r_a-(1-2k)\cos^2r_b]}, & k\in[0, \omega_7) \\
k\sin r_a\cos r_b - \sqrt{(1-k)\sin^2r_a[k+k\cos^2r_a+(1-2k)\cos^2r_b]}, & k\in(\omega_8, 1] \\
0, & k\in(\omega_7, \omega_8)
\end{cases}
\end{equation}
where
\begin{equation} \begin{aligned}
&\omega_1 =\frac{\sin^2 r_a - \sin^2 r_b + 2 - \sqrt{(\sin^2 r_a - \sin^2 r_b + 2)^2 - 4(\sin^2 r_a + \sin^2 r_b)(1 - \sin^2 r_b)}}{2(\sin^2 r_a + \sin^2 r_b)},\\
&\omega_2 =\frac{\sin^2 r_a + 3\sin^2 r_b - 2 + \sqrt{(\sin^2 r_a + 3\sin^2 r_b - 2)^2 + 4(\sin^2 r_a + \sin^2 r_b)(1 - \sin^2 r_b)}}{2(\sin^2 r_a + \sin^2 r_b)},\\
&\omega_3 =\frac{\cos^2 r_a - \cos^2 r_b + 2 - \sqrt{(\cos^2 r_a - \cos^2 r_b + 2)^2 - 4(\cos^2 r_a + \cos^2 r_b)(1 - \cos^2 r_b)}}{2(\cos^2 r_a + \cos^2 r_b)},\\
&\omega_4 = \frac{\cos^2 r_a + 3\cos^2 r_b - 2 + \sqrt{(\cos^2 r_a + 3\cos^2 r_b - 2)^2 + 4(\cos^2 r_a + \cos^2 r_b)(1 - \cos^2 r_b)}}{2(\cos^2 r_a + \cos^2 r_b)},\\
&\omega_5 =\frac{\sin^2 r_a + \sin^2 r_b + 1 - \sqrt{(\sin^2 r_a + \sin^2 r_b + 1)^2 - 4\sin^2 r_b (1 + \sin^2 r_a - \sin^2 r_b)}}{2(1 + \sin^2 r_a - \sin^2 r_b)},\\
&\omega_6 =\frac{\sin^2 r_a - 3\sin^2 r_b + 1 + \sqrt{(\sin^2 r_a - 3\sin^2 r_b + 1)^2 + 4\sin^2 r_b (1 + \sin^2 r_a - \sin^2 r_b)}}{2(1 + \sin^2 r_a - \sin^2 r_b)},\\
&\omega_7 =\frac{\cos^2 r_a + \cos^2 r_b + 1 - \sqrt{(\cos^2 r_a + \cos^2 r_b + 1)^2 - 4\cos^2 r_b (1 + \cos^2 r_a - \cos^2 r_b)}}{2(1 + \cos^2 r_a - \cos^2 r_b)},\\
&\omega_8 =\frac{\cos^2 r_a - 3\cos^2 r_b + 1 + \sqrt{(\cos^2 r_a - 3\cos^2 r_b + 1)^2 + 4\cos^2 r_b (1 + \cos^2 r_a - \cos^2 r_b)}}{2(1 + \cos^2 r_a - \cos^2 r_b)},\\
\end{aligned}\end{equation}

Furthermore, in case of $r_a=r_b=r$,  we have
\begin{equation}
C_{bf}(\rho _{A_IB_I} ) =
\begin{cases}
(1-k)\cos^2r- \sqrt{k\cos^2r[k+(2-3k)\sin^2r]}, & k\in[0, \omega_9) \\
k\cos^2r- \sqrt{(1-k)\cos^2r[1-k+(3k-1)\sin^2r]}, & k\in(\omega_{10}, 1] \\
0, & k\in(\omega_9, \omega_{10})
\end{cases}
\end{equation}

\begin{equation}
C_{bf}(\rho _{A_{II}B_{II}} )=
\begin{cases}
(1-k)\sin^2r - \sqrt{k\sin^2r[k+(2-3k)\cos^2r]}, & k\in[0, \omega_{11}) \\
k\sin^2r - \sqrt{(1-k)\sin^2r[1-k+(3k-1)\cos^2r]}, & k\in(\omega_{12}, 1] \\
0, & k\in(\omega_{11}, \omega_{12})
\end{cases}
\end{equation}

\begin{equation}
C_{bf}(\rho _{A_{I}B_{II}} )=
\begin{cases}
(1-k)\sin r\cos r - \sqrt{k\cos^2r[1-k+k\sin^2r]}, & k\in[0, \omega_{13}) \\
k\sin r\cos r - \sqrt{(1-k)\cos^2r[k+(1-k)\sin^2r]}, & k\in(\omega_{14}, 1] \\
0, & k\in(\omega_{13}, \omega_{14})
\end{cases}
\end{equation}

\begin{equation}
C_{bf}(\rho _{A_{II}B_{I}})=
\begin{cases}
(1-k)\sin r\cos r - \sqrt{k\sin^2r[1-k+k\cos^2r]}, & k\in[0, \omega_{15}) \\
k\sin r\cos r - \sqrt{(1-k)\sin^2r[k+(1-k)\cos^2r]}, & k\in(\omega_{16}, 1] \\
0, & k\in(\omega_{15}, \omega_{16})
\end{cases}
\end{equation}

where\\

\begin{equation} \begin{aligned}
& \omega_9 =\frac{1 - \sqrt{1 - 2\sin^2 r\cos^2 r}}{2\sin^2 r},\\
& \omega_{10} =\frac{2\sin^2 r - 1 + \sqrt{(2\sin^2 r - 1)^2 + 2\sin^2 r\cos^2 r}}{2\sin^2 r},\\
& \omega_{11} =\frac{1 - \sqrt{1 - 2\sin^2 r\cos^2 r}}{2\cos^2 r},\\
& \omega_{12} = \frac{2\cos^2 r - 1 + \sqrt{(2\cos^2 r - 1)^2 + 2\sin^2 r\cos^2 r}}{2\cos^2 r},\\
& \omega_{13} =\frac{1 + 2\sin^2 r - \sqrt{(1 + 2\sin^2 r)^2 - 4\sin^2 r}}{2}, \\
& \omega_{14} =\frac{1 - 2\sin^2 r + \sqrt{(1 - 2\sin^2 r)^2 + 4\sin^2 r}}{2}, \\
& \omega_{15} =\frac{1 + 2\cos^2 r - \sqrt{(1 + 2\cos^2 r)^2 - 4\cos^2 r}}{2}, \\
& \omega_{16} =\frac{1 - 2\cos^2 r + \sqrt{(1 - 2\cos^2 r)^2 + 4\cos^2 r}}{2}.\\
\end{aligned}\end{equation}
\end{widetext}

\end{document}